\documentclass[preprint,12pt]{elsarticle}

\usepackage{amssymb}

\usepackage{booktabs}
\usepackage{dcolumn}
\usepackage{array}
\usepackage{datenumber}

\setdatenumber{2013}{11}{04}

\newcommand{\rvec}[2]{\mathbf{r}_{#1}^{#2}}
\newcommand{\qvec}{\mathbf{q}}
\newcommand{\rpar}{\mathbf{x}}

\newcommand{\ellav}{\ell}
\newcommand{\Eq}[1]{Eq.~(\ref{eq:#1})}

\journal{Advances in Colloid and Interface Science}

\begin{document}

\begin{frontmatter}



\title{Disjoining Pressure and the Film-Height-Dependent Surface Tension of Thin
Liquid Films: New Insight from Capillary Wave Fluctuations}


\author{Luis G. MacDowell$^1$, Jorge Benet$^1$, Nebil A. Katcho$^2$ \\ and Jose
Mar\'{\i}a G. Palanco$^3$}

\address{$^1$Departamento de Qu\'{\i}mica F\'{\i}sica, Facultad de Ciencias
Qu\'{\i}micas, Universidad Complutense, Madrid 28040, Spain.
         \\
         $^2$LITEN, CEA-Grenoble, 17 rue des Martyrs, 38054 Grenoble Cedex 9, France.
         \\
         $^3$Departamento de Qu\'{\i}mica Aplicada, ETSI Aeronauticos, Universidad Politécnica de Madrid, 28040, Spain.
        }

\begin{abstract}
In this paper we  review simulation and experimental studies of thermal capillary wave fluctuations as an
ideal means for probing the underlying disjoining pressure and surface tensions, and more generally, fine
details of  the Interfacial Hamiltonian Model. We discuss recent simulation results that reveal 
a film--height--dependent surface tension not accounted for in the classical Interfacial Hamiltonian Model.
We show how this observation may be explained bottom--up from sound principles of statistical thermodynamics
and discuss some of its implications.
\end{abstract}


\end{frontmatter}




\section{Introduction}

As materials science and nanotechnology improve our ability to produce devices of
smaller and smaller size down to the nanoscale,  the importance of interfacial
phenomena becomes yet more relevant \cite{seemann05}. 

Indeed, current methods allow us to prepare intricate devices, which feature
grooves, channels and containers, offering the possibility to process minute
amount of liquids in a controlled manner  \cite{burns96,king03,cabanas04}.

Obviously, the operation of such devices requires  detailed understanding
of the fluid's behavior, and the size to surface ratio of the condensates
that result
makes the role of surface interactions a key issue \cite{israelachvili91}. At sub--micrometer
length scales, however, the classical surface thermodynamics of Young and Laplace 
may well not be sufficient \cite{milchev01b}. The precise
nature of the fluid--substrate interactions becomes important, and it is no longer possible
to lump all such effects into a macroscopic contact angle. 
Attempts to extend the validity of the classical thermodynamic approach are
based on the addition of line tension effects \cite{gretz66,boruvka77,churaev82,widom95},  and
provide encouraging results
\cite{bresme98,pompe00,milchev01b,wang01,pompe02,binder11b}.
However, this concept meets difficulties and controversies
\cite{degennes85,drelich96,amirfazli04,velarde11}, and is difficult
to extend beyond the study of  sessile droplets.

A well known route to study adsorption  phenomena at such length scale
refines the level of coarse--graining one step below, by describing the
properties of the adsorbed liquid in terms of a film height, $\ell$. This
provides a means to incorporate surface forces in a detailed manner
\cite{philip77,degennes85,robbins91,sharma93,starov09}, using the
celebrated Derjaguin's concept of disjoining pressure, $\Pi(\ell)$
\cite{derjaguin92,derjaguin92b}, or,
alternatively, the corresponding interface potential $g(\ell)$ \cite{dietrich88,schick90}.

Our understanding of wetting phenomena owes a great deal to such
concept. More interestingly, however, the interface potential also offers the possibility to
study the properties of inhomogeneous films, by means of a simple
phenomenological extension, known as the Interfacial Hamiltonian Model (IHM).
In this model, one defines a film profile, $\ell(\rpar)$, dictating the film
height on each point of the underlying plane. Each infinitesimal surface area
element $d\rpar$, bares a free energy $g(\ell(\rpar))$ dictated by the film
height at that point. However, since the film is inhomogeneous, an
additional contribution accounting for the increase of the liquid--vapor
interfacial area
is required. Considering both contributions, and integrating over the whole
plane of the substrate, one arrives at \cite{degennes85,starov09}:
\begin{equation}
\label{eq:ifh}
H[\ell(\rpar)] = \int  \left \{ g(\ell) + \gamma_{\infty} ( \sqrt{ 1 +
(\nabla \ell)^2 } -1 )   \right \} d \rpar
\end{equation} 
where $\gamma_{\infty}$ is the liquid--vapor surface tension, while 
the label "$\infty$" as a subindex emphasizes the fact that, for whatever
film height, we refer to the
surface tension of a film away from the influence of the disjoining pressure. i.e., 
the liquid--vapor surface tension. Since equilibrium film profiles are extrema of 
$H[\ell]$, it may be readily found that the IHM is essentially equivalent to the
augmented Young--Laplace equation that is familiar in surface science
\cite{philip77,degennes85,robbins91,sharma93,starov09}.

The importance of \Eq{ifh} should not be overlooked, as it forms the basis for
most theoretical accounts of surface phenomena, including,
the study of capillary waves \cite{buff65},
renormalization group analysis of wetting phenomena \cite{fisher85}, the
prediction of droplet profiles \cite{degennes04},
the measure of line tensions \cite{dobbs93}, the structure of adsorbed films on
patterned substrates \cite{bauer99b}, and the dynamics of dewetting \cite{vrij66}.

Despite its theoretical importance, it has been argued for already some time that  the
IHM cannot be derived bottom--up from a microscopic Hamiltonian of finer
coarse--graining level \cite{safran94}.
This issue has received a great deal of attention in the context of adsorbed
fluids subject to a short--range wall--fluid potential. This system exhibits
a critical wetting transition, the liquid film can grow almost unbound, and the 
interfacial fluctuations become increasingly large \cite{schick92}. As a result,
the wetting behavior cannot be accounted properly by the mean field interface
potential, $g(\ell)$, but rather, must be described by suitable renormalization of
\Eq{ifh}.
Conflicting results of the theoretical analysis \cite{fisher85} with simulations
\cite{binder88}, motivated
a critical assessment on the foundation of IHM \cite{jin93,fisher94,parry06}. 
Fisher and Jin attempted to derive \Eq{ifh} using a Landau--Ginzburg--Wilson
Hamiltonian, and argued that this is possible provided one replaces
$\gamma_{\infty}$ by a film--thick dependent surface tension,
$\gamma(\ell)$ which approaches $\gamma_{\infty}$ exponentially fast
\cite{jin93,fisher94}. 
However, further studies by Parry and collaborators have shown that IHM is 
actually a nonlocal functional which does generally not satisfy \Eq{ifh}, except
for some simple situations \cite{parry06,parry07}.

Unfortunately, these studies are  limited to the special case of short--range forces, 
which are only found in nature under exceptional circumstances, as they correspond to an effectively
vanishing Hamaker constant \cite{ragil96,shahidzadeh98}. The  more
relevant case of fluids in the presence of van der Waals interactions,
 has apparently received much less attention
\cite{bernardino08,bernardino09,macdowell13}, possibly because the long--range interactions
inhibit fluctuations and do not warrant a renormalization analysis.

However, the problem remains an issue of great importance for the study of
inhomogeneous films under confinement--condensed sessile drops, fluids
adsorbed in grooves, and other condensed structures--irrespective of the presence of critical
fluctuations!

In fact, thin adsorbed films subject to van der Waals forces still exhibit
thermal surface fluctuations of amplitudes as large as the $mm$ scale that are known under the name of capillary waves
\cite{buff65,rowlinson82b}. The study of this, less exquisite fluctuations actually can
convey not only a great deal of information on the underlying surface forces
\cite{tidswell91,doerr99,mora03},
but is actually also a stringent test of the Interface Hamiltonian Model itself
\cite{macdowell13,pang11,fernandez12}.

In this paper, we will review studies of the capillary wave fluctuations of
adsorbed films performed over the last years, and describe recent
findings which shed some light on the conjectured dependence of the surface
tension with film height \cite{macdowell11,gregorio12,macdowell13}.

In the next section, we will give an overview of well known liquid--state
theories for the description of density profiles of planar adsorbed films. Since
these theories are of mean field type, they lead to structural properties which
are {\em intrinsic} to the fluid--substrate pair considered, and do not depend
on other external considerations such as the system size. In section 3, we give a
brief overview of classical capillary wave theory and show how it allows to
probe the interfacial structure of films as well as to validate  the Interfacial
Hamiltonian Model. We illustrate the classical predictions with a number of experiments 
and computer studies, and show how the capillary fluctuations renormalize the 
intrinsic density profiles, which actually become system size dependent and are
therefore not intrinsic properties of the fluid--substrate pair.
In Section 4 we  describe computer simulation techniques for the study of capillary
wave fluctuations, and  discuss how very recent simulation evidence has gathered
that calls for an improved interfacial Hamiltonian model. This problem is
reviewed in section 5, where  the results of section 2 are applied
in order to derive an interfacial Hamiltonian bottom--up, for fluid films
subject to  surface forces decaying well beyond the bulk liquid correlation
length, as is usually the case in real systems exhibiting dispersion forces.
Finally,  section 6 summarizes the outcome of the study and discusses some of
its implications.

\section{Liquid state theory of adsorbed fluids}

\label{sec:dft}

In modern liquid--state theory, the study of interfaces is formulated in terms
of free energy functionals of the number density, $\rho(\rvec{}{})$
\cite{evans92}.
Paralleling the expression of the Helmholtz free energy of a volumetric system,
i.e., $F=N k_BT(\ln\Lambda^3\rho - 1) + F_{\rm ex}$, which includes ideal and excess
contributions, one writes, for the inhomogeneous system, the following density
functional:
\begin{equation}\label{eq:ffunc}
  F[\rho(\rvec{}{})] = k_B T \int \rho(\rvec{}{})\left ( \ln \Lambda^3 \rho(\rvec{}{}) - 1 \right ) d \rvec{}{}
  + F_{\rm ex}[\rho(\rvec{}{})]
\end{equation} 
where $N$ is the number of molecules, $k_B$ is Boltzmann's constant,  $T$ is
the absolute temperature and $\Lambda$ is the thermal de Broglie wavelength, 
while $F_{\rm ex}$ is a highly non--trivial functional
incorporating all unknown multibody correlations. 

In practice, it is more convenient to
relax the constraint over fixed number of particles that is appropriate for
Helmholtz free energies, and consider a system with fixed bulk chemical
potential $\mu_{\infty}$. This is achieved by introducing the grand free energy
$\Omega$,  a new functional of the density  which can be obtained from
$F$  by  Legendre transformation, $F - \mu_{\infty} N \to \Omega$. 
\begin{equation}\label{eq:omegafunc}
   \Omega[\rho(\rvec{}{})] = F[\rho(\rvec{}{})] + \int \rho(\rvec{}{}) (  V(\rvec{}{}) -  \mu_{\infty} )
   d \rvec{}{}
\end{equation}
where we have also included here $V(\rvec{}{})$, an external field that will
usually be the responsible for creating the inhomogeneity under study. For an
adsorbed fluid, $V(\rvec{}{})$ may be the van der Waals long range potential
mimicking the interactions with the substrate; for a free fluid--fluid interface 
it may be the potential energy felt by an atom due to gravity.  

Within mean--field theory, we expect that the equilibrium average profile  is
that which minimizes $\Omega$, subject to the constraints of
constant volume, temperature and chemical potential: 
\begin{equation} 
\frac{\delta \Omega}{\delta \rho(\rvec{}{})} = 0
\end{equation}

Performing the functional minimization of \Eq{omegafunc}, together with
\Eq{ffunc}, we obtain:
\begin{equation}\label{eq:GS}
   \rho(\rvec{}{}) = \rho_{\infty}\exp\{-\beta V(\rvec{}{}) +
               C^{(1)}(\rvec{}{}) + \beta\mu_{\rm ex} \}
\end{equation}
where $\beta=1/k_BT$, $\rho_{\infty}$ is the bulk density at the imposed chemical
potential, $\mu_{\rm ex}$ is the corresponding excess chemical
potential and $C^{(1)}(\rvec{}{})$ is the so called singlet direct
correlation function:
\begin{equation} \label{eq:C1}
  C^{(1)}(\rvec{}{}) = - \frac{\delta \beta F_{\rm ex}[\rho(\rvec{}{})]}{\delta
  \rho(\rvec{}{})}
\end{equation}
The above equation is the first member of a hierarchy defining
direct correlation functions of arbitrary order \cite{rowlinson82b,henderson92}. The next
member of the series provides the direct pair correlation function as:
\begin{equation}
 C^{(2)}(\rvec{}{},\rvec{}{'}) = \frac{\delta
 C^{(1)}(\rvec{}{};[\rho])}{\delta \rho(\rvec{}{'})}
\end{equation}
By integrating the above equation from some density reference
profile, $\rho_0(\rvec{}{})$, to the actual density profile,
we obtain:
\begin{equation}
     C^{(1)}(\rvec{}{};[\rho]) = C^{(1)}(\rvec{}{};[\rho_0])
     + \int \int
      C^{(2)}(\rvec{}{},\rvec{}{'};[\rho])\,
      \delta\rho\, d  \rvec{}{'}
\end{equation}
This equation is formally exact but of little use, since we
ignore the exact form of both $C^{(1)}(\rvec{}{};[\rho_0])$ and
$C^{(2)}(\rvec{}{},\rvec{}{'};[\rho])$. We can however,
consider a flat reference profile, such that
$\rho_0(\rvec{}{})=\rho_{\infty}$, and further assume that the direct
pair correlation function does not depend significantly on
deviations of $\rho(\rvec{}{})$ away from the reference density $\rho_0$. With
these
approximations we obtain an asymptotic density expansion:
\begin{equation}\label{eq:LDE}
     C^{(1)}(\rvec{}{};[\rho]) =
     C^{(1)}(\rho_{\infty})
     + \int  \Delta\rho(\rvec{}{'})\,
      C^{(2)}(\rvec{}{},\rvec{}{'};\rho_{\infty}) d  \rvec{}{'}
\end{equation}
where $\Delta\rho(\rvec{}{})=\rho(\rvec{}{}) - \rho_{\infty}$, while
the unknown singlet correlation function is now expressed
in terms of singlet and pair correlation functions of a
homogeneous fluid with asymptotic density $\rho_{\infty}$.

This is a convenient result, because much is known about the direct correlation
function of bulk simple fluids \cite{hansen86,mcquarrie76}. One could thus employ such
knowledge to calculate $ C^{(1)}(\rvec{}{};[\rho])$ accurately and exploit
\Eq{GS} and \Eq{LDE}
to predict the density profile. Unfortunately such a program can only be carried
out with heavy numerical calculations \cite{tang04,tang05,tang07}. In order to obtain tractable
expressions, it is necessary to get rid of the nonlocal integral by performing a
gradient expansion of the density difference $\Delta\rho(\rvec{}{'})$
about $\Delta\rho(\rvec{}{ })$ to second order. Considering that the
bulk direct correlation function of an atomic fluid is an even function of $|\rvec{}{'}-\rvec{}{
}|$, we find that odd terms in the expansion vanish, and get \cite{barker82}:
\begin{equation}\label{eq:LDCH}
     C^{(1)}(\rvec{}{};[\rho]) =
     C^{(1)}(\rho_{\infty})
     + \left
     (\frac{1}{\rho_{\infty}}-\frac{\beta}{\kappa_{\infty}\rho_{\infty}^2} \right )
     \Delta\rho(\rvec{}{ }) + C_{\infty}\nabla^2
     \Delta \rho(\rvec{}{})
\end{equation}
where the coefficient linear in $\Delta\rho(\rvec{}{ })$ is the zero order moment of the
direct pair correlation function and may be related to the bulk compressibility,
$\kappa_{\infty}$, via the Ornstein--Zernike equation \cite{rowlinson82b}:
\begin{equation}\label{eq:mom0}
   \int C^{(2)}(\rvec{}{};\rho_{\infty}) d  \rvec{}{} =
\frac{1}{\rho_{\infty}}- \frac{\beta}{\kappa_{\infty}\rho_{\infty}^2}
\end{equation}
and $C_{\infty}$ is  the second moment of the direct pair correlation function:
\begin{equation}\label{eq:mom2}
     C_{\infty} = \frac{1}{6} \int \rvec{}{2} C^{(2)}(\rvec{}{};\rho_{\infty}) d  \rvec{}{}
\end{equation}
If we now substitute \Eq{LDCH} into \Eq{GS} and linearize the exponential
term, we find that $\Delta \rho(\rvec{}{})$
is determined by the following second order partial differential equation:
\begin{equation}\label{eq:helmholtz}
 \nabla^2 \Delta \rho(\rvec{}{}) -
 b_{\infty}^2 \Delta \rho(\rvec{}{}) = \frac{\beta}{ C_{\infty}} V(\rvec{}{})
\end{equation}
where $b_{\infty}=\xi_{\infty}^{-1}$, while $\xi_{\infty}$ is the bulk correlation length, given by:
\begin{equation}
  \xi_{\infty}^2 = k_B T C_{\infty}\kappa_{\infty}\rho_{\infty}^2
\end{equation}  
Essentially, \Eq{helmholtz} corresponds to a square--gradient theory for the
Helmholtz free energy functional, with a parabolic approximation for the local free
energy (see below). The advantage of the systematic derivation from first principles is that
a deeper insight on the nature of the square--gradient coefficient is obtained.

Unfortunately, this equation has one very important limitation that may have
been overlooked: it relies on a gradient expansion of the density perturbations,
$\Delta\rho(\rvec{}{})$. 
This implies that the coefficients of the successive
derivatives are moments of the direct correlation function (e.g. as is the case
of the coefficient of $\nabla^2 \Delta\rho$, c.f. \Eq{mom2} ).  The direct correlation
function itself is known to decay as the underlying pair potential so that
the higher order moments can only converge if the fluid pair potential is short
range, i.e., is either truncated at a finite value or decays exponentially fast.
These considerations imply that the results obtained so far are strictly valid 
only for short--range fluids with exponential decay of the pair interactions
\cite{evans92}.
Later on we will discuss at length the significance of these limitations (c.f.
section \ref{sec:lvs}).

In what follows, we will exploit the above result in order to study density
profiles of the liquid--vapor and wall--liquid interfaces.  
For such systems, the average density profile depends only
on the perpendicular distance to the interface, $z$, so that the Helmholtz
equation becomes a simple linear ordinary differential equation:
\begin{equation}\label{eq:EDO}
 \frac{d^2 \Delta \rho(z)}{d z^2} - b_{\infty}^2 \Delta\rho(z) =  
        \frac{\beta}{ C_{\infty}} V(z)
\end{equation} 
Later on, we will see that the approach starting from \Eq{LDE} can also be extended to study
oscillatory profiles of fluids adsorbed on a wall.

\subsection{Liquid--Vapor Interface}

\label{sec:lvi}

\begin{figure}[t]
\centering
\includegraphics[clip,scale=0.30,trim=0mm 5mm 0mm 0mm,
angle=270]{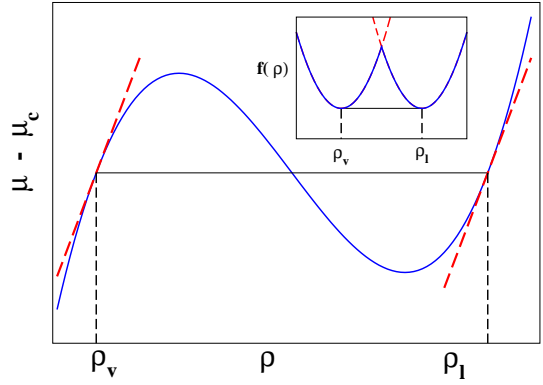}
\caption{
Sketch of the double parabola approximation for the chemical potential isotherm.
Full lines depict the full van der Waals loop in the $\mu-\rho$ plane, and the
dashed lines illustrate the linearisation that is performed about the
coexistence vapor and liquid densities. The linear extrapolation of the equation
of state results from differentiation of a double parabolic model for the free
energy (inset).
\label{fig:eosdp}
}
\end{figure}

Let us now consider the inhomogeneous density profile that results when a
homogeneous fluid phase separates at zero field, such that $V(z)=0$.
In principle, it is impossible to study a liquid--vapor interface from the result
of \Eq{EDO}. The reason is that it is essentially a local expansion about a
reference bulk density, say the vapor density, and hence, cannot possibly carry any
information about the liquid phase. In practice, however, one can exploit this
result to study perturbations of the liquid and vapor branches of the density
profile independently and get a full liquid--vapor density profile by matching the
separate pieces.  This approximation constitutes the {\em double--parabola
model} of
interfaces \cite{jin93}. The   name stems from the parabolic approximation 
about the vapor and liquid minima of the local Helmholtz free energy that is
implied. This becomes more clear if we consider the well known
square--gradient functional of inhomogeneous fluids \cite{cahn58}:
\begin{equation}\label{eq:Fsgt}
       \beta F[\rho(\rvec{}{})] = \int \left \{ \beta f(\rho(\rvec{}{})) + \frac{1}{2}\,
	 C_{\infty}
	    \left [ \nabla \rho(\rvec{}{}) \right ]^2
	    \right \} d  \rvec{}{}
\end{equation}
where the local free energy is given by $f(\rho)=\rho \mu(\rho) - p(\rho)$ and
$p(\rho)$ is the pressure.
Substitution of this square--gradient functional into \Eq{omegafunc}, followed by
extremalisation, yields:
\begin{equation}\label{eq:Fsgte}
		    C_{\infty} \nabla^2 \Delta\rho(\rvec{}{}) - \beta [\mu(\rho) -
		    \mu_{\infty} ] = \beta V(\rvec{}{}) 
\end{equation}
This equation cannot be solved analytically for a general isotherm $\mu(\rho)$,
but can be related to \Eq{EDO} upon linearisation of the chemical potential
about the coexistence value, whereby $\mu(\rho)-\mu_{\infty}$ becomes simply
$(\rho-\rho_{\infty})/(\rho_{\infty}^2 \kappa_{\infty})$, and \Eq{Fsgte} then
immediately transforms into \Eq{helmholtz}. Linearising the isotherm about the
coexistence liquid and vapor densities, and solving for each branch separately
constitutes the double parabola approximation. 
The linearised chemical potential isotherm results from derivation of a parabolic 
Helmholtz free energy centered about the coexistence density, 
thus explaining the name of the model (c.f. Fig.\ref{fig:eosdp}).

Consider the liquid--vapor interface is located at  $z=\ellav$, with the
asymptotic liquid phase of density $\rho_l$ to the left ($z<\ellav$), and the asymptotic 
vapor phase of density $\rho_v$ to the right of $\ellav$ ($z>\ellav$). With this boundary
conditions so defined, we can solve  \Eq{EDO} for each branch 
separately, obtaining a piecewise solution of the form: 
\begin{equation}\label{eq:piecewise}
\rho_{\rm lv}(z) =
 \left \{
 \begin{array}{cc}
   \rho_l + A_l e^{b_l z} & z < \ellav  \\
   \\
   \rho_v + A_v e^{-b_v z} & z > \ellav
 \end{array}
\right .
\end{equation} 
In order to solve for the integration constants,  $A_l$ and $A_v$, two
possible extra boundary conditions come to mind.  The first is the {\em crossing
criterion}
\cite{jin93,fisher94}, which
requires the continuity of the piecewise function, \Eq{piecewise} at $z=\ellav$, 
and defines $\ellav$ such that the density at that point is precisely some chosen
value, say, $\rho_{1/2}$, which is, most naturally, but not necessarily
 equal to the average $(\rho_l + \rho_v)/2$: 
\begin{equation}\label{eq:crossing}
 \left \{
 \begin{array}{ccc}
   \lim_{z\to\ellav^-} \rho_{\rm lv}(z)  & = &   
                \lim_{z\to\ellav^+} \rho_{\rm lv}(z)  \\
   \\
   \rho_{\rm lv}(z=\ellav) & = &  \rho_{1/2}   \\
 \end{array}
\right .
\end{equation} 
The crossing criterion provides a set of two linear equations that can be easily
solved for $A_l$ and $A_v$, and leads to the following result for the piecewise
liquid--vapor density profile \cite{jin93}:
\begin{equation}\label{eq:dplvcc}
\rho_{\rm lv}(z) =
 \left \{
 \begin{array}{cc}
   \rho_l +  (\rho_{1/2} - \rho_l) 
                      e^{b_l ( z - \ellav)} & z < \ellav  \\ 
   \\
   \rho_v +  (\rho_{1/2} - \rho_v) 
                      e^{-b_v ( z - \ellav)} & z > \ellav 
 \end{array}
\right .
\end{equation} 
This model has proved very convenient, as it  provides analytic
results for the density profiles and free energies of interfaces perturbed by capillary waves
\cite{jin93,fisher94,parry04,parry06,bernardino08,bernardino09}. Although we have cast it here in a form that
accounts for the asymmetry of the vapor and liquid phases, most usually one
assumes a symmetric fluid, hence $b_l=b_v$.

As just mentioned, the crossing criterion would seem to account for the
asymmetry of the vapor and liquid phases. However, the first derivative of the density
profile becomes discontinuous whenever $b_l\neq b_v$. In order to remedy this
problem, it is possible to introduce a {\em smooth matching criterion}  by  requiring
continuity of both the density and its first derivative at $z=\ellav$ i.e.,
\begin{equation}\label{eq:dplvmc}
 \left \{
 \begin{array}{ccc}
   \lim_{z\to\ellav^-} \rho_{\rm lv}(z) =   \lim_{z\to\ellav^+} \rho_{\rm lv}(z)  \\
   \\
   \lim_{z\to\ellav^-} \frac{d \rho_{\rm lv}(z)}{d z} =   \lim_{z\to\ellav^+} 
                \frac{d \rho_{\rm lv}(z)}{d z}  \\
 \end{array}
\right .
\end{equation} 
Solving the matching conditions for  $A_l$ and $A_v$, now yields:
\begin{equation}\label{eq:dplv}
\rho_{\rm lv}(z) =
 \left \{
 \begin{array}{cc}
   \rho_l - \frac{b_v}{b_v + b_l} (\rho_l - \rho_v) 
                      e^{b_l ( z - \ellav)} & z < \ellav  \\ 
   \\
   \rho_v +  \frac{b_l}{b_v + b_l} (\rho_l - \rho_v) 
                      e^{-b_v ( z - \ellav)} & z > \ellav 
 \end{array}
\right .
\end{equation} 
Despite its simplicity, the model incorporates naturally the asymmetry of the
liquid and vapor phases, remains continuous up to the first derivative, and is able to provide semi--quantitative results for
the density profiles and surface tensions of simple fluids \cite{palanco13}.
Furthermore, the model may be extended to study spherical
interfaces, also providing analytical results for density profiles and
nucleation energies \cite{iwamatsu93,iwamatsu94,bykov99b,palanco13}.

Such analytical results cannot be obtained otherwise except for very few
selected toy models (e.g.: \cite{hemingway81,rowlinson82b}).

\subsection{Wall--liquid Interface}

\label{sec:wfi}

Simple atomic liquids close to a rigid
substrate  exhibit a stratified structure that results
from packing effects of the dense phase. Such behavior is well known from both theoretical
calculations and atomic force microscopy experiments
\cite{ebner77,horn81,chernov88,tarazona83,tarazona85,henderson05,klapp08}.

The model for fluid interfaces discussed in the previous section would seem not adequate to
describe this behavior, since it may be interpreted as a plain
squared gradient theory solved piecewise.
With this perspective, one  can only  expect it to
provide adequate results for smoothly varying density perturbations. However, it
is possible to exploit the explicit connection with the direct pair correlation
function embodied in \Eq{LDCH}--\Eq{mom2} in order to provide a qualitative explanation for the oscillatory
behavior found in experiments. First, notice that the coefficients of \Eq{LDCH} are
actually zero and second moments of the direct correlation function. Whence, 
they can also be interpreted as their zero wave--vector Fourier transforms. 
Taking this into account, it becomes apparent
that the theory formulated previously is adequate to study perturbations of
long wavelength only. 

Molecular fluids at high density usually exhibit a maximum
of the structure factor, $S(k)$ at finite wave--vector, $k_{\rm o}$. Such a maximum is
indicative of strong structural correlations of wavelength $\lambda=2\pi/k_o$. Accordingly,
it seems natural to  particularize the study of density 
fluctuations  to the form
$\Delta\rho(\rvec{}{})=a(\rvec{}{}) e^{\pm i {\bf k}_{\rm o}\cdot \rvec{}{}}$, where the
second factor of the right hand side now imposes correlations of the adequate
wavelength, while the first factor, $a(\rvec{}{})$ describes the corresponding
amplitudes. In the regime of linear response, it is these amplitudes that 
should vary smoothly, rather than the whole density wave $\Delta\rho(\rvec{}{})$.
Therefore, one can perform  a  gradient expansion of $a(\rvec{}{'})$ about
$a(\rvec{}{})$, similar to that performed previously for $\Delta\rho(\rvec{}{'})$
about $\Delta\rho(\rvec{}{})$. 
After insertion of the expansion into \Eq{LDE}, 
followed by substitution in the linearized form of \Eq{GS},
we obtain a Helmholtz equation for the amplitudes rather than for the densities \cite{tarazona84}:
\begin{equation}\label{eq:helmholtza}
 \nabla^2 a(\rvec{}{}) -
 b^2_{\rm o} a(\rvec{}{}) = 0
\end{equation}
where now, the coefficient $b_{\rm o}=1/\xi_o$, while $\xi_o^2=k_BTC_o\kappa_o\rho_{\infty}^2$ is given in terms of 
a generalized wave--vector dependent compressibility (c.f. \Eq{mom0}):
\begin{equation}\label{eq:mom0k}
   \int C^{(2)}(\rvec{}{}) e^{i {\bf k}_o \cdot \rvec{}{}} d  \rvec{}{} =
\frac{1}{\rho_{\infty}}
   -\frac{\beta}{\kappa_o\rho_{\infty}^2}
\end{equation}
and $C_o$ is a Fourier transform of the  direct pair correlation function's second
moment:
\begin{equation}\label{eq:mom2k}
     C_o = \frac{1}{6} \int \rvec{}{2} C^{(2)}(\rvec{}{}) e^{i {\bf k}_o
     \cdot \rvec{}{}} d  \rvec{}{}
\end{equation}

Let us consider the solution of \Eq{helmholtza} for the simple case of a
``contact'' potential of delta--Dirac form whose only role is to impose a
boundary condition for the density of the film precisely at the wall contact,
$z=d_w$.
The solution of this equation proceeds then exactly as for \Eq{helmholtz}, and
yields for the wall--liquid density profile the following result:
\begin{equation}
  \rho_{wl}(z) = \rho_l ( 1 + A_w cos(k_{\rm o} z + \theta_w) e^{-b_{\rm o} z} )
\end{equation} 
where $A_w$ is the amplitude of the density wave imposed by the contact wall
potential and $\theta_w$ is the
phase. An interesting point which is worth stressing is that both  $k_o$ and
$b_{\rm o}$ are structural properties of the bulk liquid. Only the amplitude
$A_w$ and the phase $\theta_w$ actually depend on details of the wall--fluid
interactions.  

\begin{figure}[t]
\centering
\includegraphics[clip,scale=0.30,trim=0mm 5mm 0mm 0mm,
angle=270]{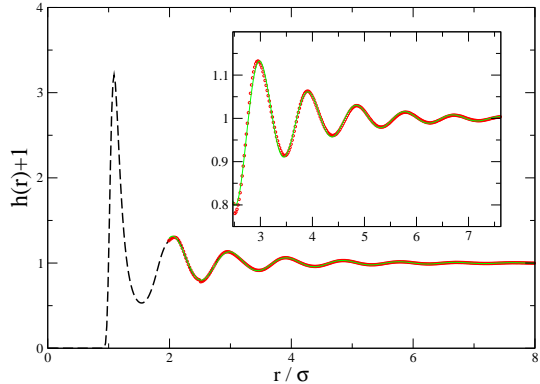}
\caption{Bulk total correlation function of a Lennard--Jones fluid close
to its triple point.
The dashed line corresponds to simulation results, and the
full line is the result of a fit to 
$r h(r) = A_o cos(k_o r + \theta_f) exp(-b_o r)$ (c.f. \Eq{hbulkasymp}). 
Symbols indicate simulation results in the range where the
fit is performed. The inset shows a detailed view
 (c.f.  Ref.\cite{macdowell13,gregorio12} and Sec.~\ref{sec:simnew} for further details on 
the model and simulations). 
\label{fig:gr_fit}
}
\end{figure}

The result shown here is actually a particular case of a more general theory
relating the density profile of an adsorbed fluid with its bulk structural
properties \cite{evans94,henderson94,evans09}. A study of the Ornstein--Zernike equation shows that, quite
generally, the total pair correlation function, $h(r)$ of an isotropic fluid is
given by:
\begin{equation}
 r h(r) = \sum_j A_j e^{i k_j r}
\end{equation} 
where the sum runs over the {\em poles} of the structure factor,
i.e., the set of complex wave--vectors satisfying $\rho C^{(2)}(k_j) - 1=0$
\cite{evans94}.
If, on the other hand, one considers the wall--liquid total correlation function,
$h_{\rm wl}(z)$, the Ornstein--Zernike equation dictates rather that:
\begin{equation}
  h_{\rm wl}(z) = \sum_j B_j e^{i k_j z}
\end{equation} 
where the sum runs over exactly the same set of wave--vectors than before, and
only the coefficients $B_j$ are actually dependent on the wall--fluid substrate.
A lucky coincidence is that only the first few leading order terms in this
expansion are necessary to obtain a very precise description of the fluid
structure. Particularly, 
for fluids with short range forces, a formal study reveals that the two longest
range wave--vectors, $k_j$ are,
i) a purely imaginary pole,  leading to pure exponential
decay, and ii) a conjugate--pair of complex poles, leading to damped oscillatory decay.

Therefore, the long range decay of the bulk pair correlation function is of the form:
\begin{equation}\label{eq:hbulkasymp}
  r h(r) = A_e e^{-b_e r} + A_o cos(k_o r + \theta_f) e^{-b_o r}
\end{equation}
while that of the wall--liquid pair correlation function is given by:
\begin{equation}\label{eq:hwf}
h_{wl}(z)  =  B_e e^{-b_e z} +  B_o cos(k_o z + \theta_w) e^{-b_o z} 
\end{equation} 
For high temperature, $b_o > b_e$, and the long--range decay is purely
monotonic. At lower temperatures, however, the contrary holds and the
long--range decay becomes damped--oscillatory. These two regimes are separated
in the temperature--density plane of the phase diagram as a line $T_{\rm
fw}(\rho)$ that is known as the Fisher--Widom line \cite{evans93b}. Actually, at temperatures close
to the triple--point, the monotonic contribution is of such short range that
only one damped--oscillatory term serves to precisely describe the pair
correlation function beyond the first maximum.

\begin{figure}[t]
\centering
\includegraphics[clip,scale=0.30,trim=0mm 5mm 0mm 0mm,
angle=270]{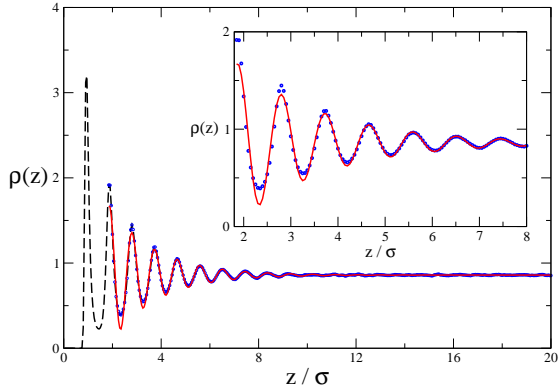}
\caption{Density profile at the wall--liquid interface of a Lennard--Jones fluid
close to its triple point.  The dashed line corresponds to simulation results.
The full line is the result of a fit to $B_o cos(k_o z + \theta_w) exp(-b_o z)$ 
(c.f. \Eq{hwf}) with $k_o$ and $b_o$ obtained
from the bulk correlation function and only $B_o$ and $\theta_w$ as fitting
parameters. Symbols indicate simulation results in the range where the
fit is performed. The inset shows a detailed view
(c.f. Ref.\cite{macdowell13,gregorio12} and Sec.~\ref{sec:simnew} for further details on the
model and simulations). 
\label{fig:rho_fit}
}
\end{figure}

The accuracy of this prediction has been assessed in Density Functional Studies
\cite{henderson94},
as well as experimentally \cite{klapp08,klapp08b}. As an example,
Fig.\ref{fig:gr_fit} shows the simulated bulk total correlation function of a
Lennard--Jones model of Argon close to its triple point. Clearly, a strong
oscillatory behavior is visible, but all of the correlation function may be
accurately described beyond two molecular diameters, $\sigma$, with a single damped
oscillatory term. Simulating now liquid Argon at the same thermodynamic
conditions but adsorbed to an attracting wall, provides the density profile
given in Fig.\ref{fig:rho_fit}. Using only the damped oscillatory term of \Eq{hwf}, with
$k_o$ and $b_o$ from the fit to the bulk correlation function, and only the
amplitude $B_o$ and phase $\theta_w$ as new fitting parameters, provides again an
excellent description beyond two molecular diameters. Such a particularly simple
behavior is a result of the low temperatures considered. At higher temperature,
at least the leading order purely exponential contribution needs to be added.

It should be stressed, however, that \Eq{hbulkasymp}--\Eq{hwf} are only appropriate for
fluids with short--range forces. This is almost always the case in simulation
studies, since the dispersion tail $r^{-6}$ is in practice, truncated beyond
some reasonable value.
Taking van der Waals contributions for the fluid--fluid pair potential into account
makes the formal analysis become far more difficult, but it is expected that the
gross features described here will still hold \cite{evans09}. For example, it is
well known that the tails of the liquid--vapor density profile of a long--range
fluid with interactions of the form  $r^{-6}$ will decay as $z^{-3}$, instead of
exponentially \cite{barker82}, but these finer details
need not concern us here. Surprisingly, even
van der Waals wall--liquid interactions of range $z^{-3}$
actually have a negligible effect on the structure of the density profile. This
can be assessed by exploiting yet once more \Eq{helmholtz}, as a means to measure the density
fluctuation $\Delta\rho(z)$ that results from a long   range perturbation
$V(z)\propto z^{-3}$. Noticing that \Eq{piecewise} already provides the homogeneous solution for
\Eq{EDO}, we seek for a particular solution of  the form:
\begin{equation}
  \Delta\rho(z) = \sum_{i=0}^{\infty} a_i\, V^{(i)}(z)
\end{equation} 
where $a_i$ are undetermined coefficients and $V^{(i)}$ stands for the $i$th
derivative of $V(z)$. Note that the particular solution suggested is actually
valid for algebraically decaying potentials. For an exponential decay, the first term of
the series would suffice. Now, substitution of the trial form into \Eq{EDO}, followed
by identification of the coefficients, yields:
\begin{equation}\label{eq:solpar}
 \Delta\rho(z) = - \frac{\beta}{C_{\infty} b_{\infty}^2} \sum_{i=0}^{\infty}
 \left ( \frac{1}{b_{\infty }^2} \right )^i V(z)^{(2i)} 
\end{equation} 
Considering that, by virtue of \Eq{mom0}, the prefactor is essentially dictated by the
fluids susceptibility, $d \rho/d \mu$, if follows that the density profile of
incompressible fluids will be hardly affected by the long--range substrate
potential. In order to grasp more transparently the significance of the above
equation, it is now convenient to perform a resummation of the linearized result. 
Employing \Eq{solpar} in order to evaluate \Eq{LDCH}, followed by substitution of the result 
into \Eq{GS} and neglect of the higher order terms, yields the following 
more familiar equation:
\begin{equation}\label{eq:generalbar}
  \rho(z) = \rho_{\infty} e^{ - \beta  \kappa_{\rm rel} V(z) }
\end{equation} 
where $\kappa_{\rm rel}=\kappa_{\infty}/\kappa_{ig}$ is the ratio of bulk to ideal gas compressibilities.
This result is thus  essentially a generalization of the
barometric law for dense fluids. For small densities, $\kappa_{\rm rel}=1$, and
\Eq{generalbar} becomes the ideal gas distribution under an external field. Close to the triple point of Argon, 
however, the ratio 
$\kappa_{\rm rel}$ is of the order $10^{-2}$, and the density profile is then
hardly perturbed except for the immediate vicinity of the substrate. This form
of the asymptotic behavior of the density profile can also be obtained from an
analysis of the Ornstein--Zernike equation \cite{evans86b,henderson94}.

\subsection{Adsorbed Films}

\label{sec:afi}

Previously, we have obtained analytic results for the density profile of a
liquid--vapor and a wall--liquid interface. 
We are now in a good position to consider the density profile of an adsorbed
film of finite thickness $\ellav$, which one expects, should exhibit structural
properties that are similar to 
those of the liquid--vapor interface in the neighborhood of $\ellav$ and
similar to those of the wall--liquid interface as one approaches the substrate.

In principle, one could employ the double parabola
model of section \ref{sec:wfi} in order to obtain the full density profile of such an adsorbed film.
This can be achieved by adding an extra exponential tail $A_w e^{-b_l z}$ into
the trial solution for the liquid branch, and solving for the constant $A_w$
with a new boundary condition at the wall. This leads to a smooth density
profile which may exhibit either an enhanced or  depleted contact
density at the wall depending on the boundary condition that is imposed
\cite{jin93}.
If one is willing to describe the  oscillations that propagate from the wall, it
suffices to seek for solutions of the liquid branch where the new monotonic
exponential tail is replaced with an oscillatory tail $A_w e^{-(b_o + i k_o)z}$.

In practice, however, we find that retaining the form of the wall--liquid and
liquid--vapor profiles, and superimposing the former on the latter actually
works much better. In this superposition approximation, we write for the film
profile:
\begin{equation}\label{eq:superpos0}
\rho(z;\ellav) = [ 1 + h_{\rm wl}(z) ]\, \rho_{\rm lv}(z;\ellav)
\end{equation} 
where $h_{\rm wl}(z)$ has the form of \Eq{hwf}, while $\rho_{\rm lv}(z;\ellav)$  is
the liquid--vapor density profile in the double parabola approximation, \Eq{piecewise}. The integration constants $A_l$
and $A_v$ may be readily calculated within the crossing criterion, providing the
following piecewise profile for the adsorbed film:
\begin{equation}\label{eq:superpos}
\rho(z;\ellav) = [ 1 + h_{\rm wl}(z) ] \times
 \left \{
 \begin{array}{cc}
   \rho_l + \frac{\rho_{1/2} - \rho_l [ 1 + h_{\rm wl}(\ellav) ]}{1 + h_{\rm
   wl}(\ellav)} e^{b_l ( z - \ellav)} & z < \ellav  \\ 
   \\
   \rho_v + \frac{\rho_{1/2} - \rho_v [1 +   h_{\rm wl}(\ellav)]}{1 + h_{\rm
   wl}(\ellav)} e^{-b_v ( z - \ellav)} & z > \ellav  \\ 
 \end{array}
\right .
\end{equation} 
As is the case of the crossing criterion applied to the liquid--vapor interface,
the resulting profile is continuous at $z=\ellav$, but its derivative is not. In
principle, one could also apply the smooth matching criterion here, but the equations
that result are far too lengthy.

\begin{figure}[t]
\centering
\includegraphics[clip,scale=0.30,trim=0mm 5mm 0mm 0mm,
angle=270]{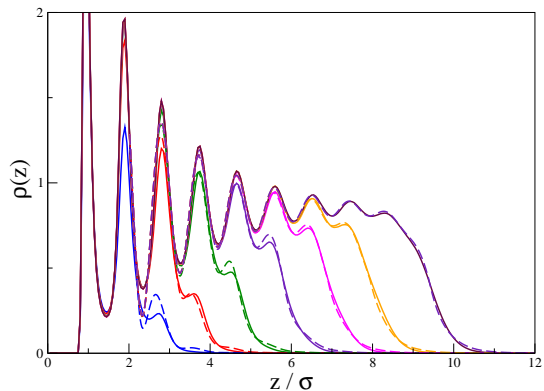}
\caption{Density profiles of adsorbed films for the Lennard--Jones fluid
close to its triple point. Full lines are  simulation results for films of
thickness $\ellav$ (from left to right) 1.9, 3.0, 4.1, 5.3, 6.5, 7.3 and 8.5
molecular diameters, $\sigma$. Dashed lines are predictions from the superposition model,
\Eq{superpos}.  Note that the discontinuity in the first derivative of the density profile is
hardly visible (results adapted from Ref.\cite{macdowell13, benet11}
correspond to the same model and temperature as the two
preceding figures).
\label{fig:profiles}
}
\end{figure}

The fact is that, already at the level of the crossing approximation,
\Eq{superpos} predicts simulated density profiles with surprising accuracy. Figure
\ref{fig:profiles}
shows a series of density profiles for films of a Lennard--Jones model of Argon 
on an adsorbing substrate in the neighborhood of the triple point. The
structural parameters  for  $h_{\rm wl}(z)$ are obtained from  results of the
wall--liquid interface described previously in Fig.\ref{fig:rho_fit}, while the
inverse correlation lengths $b_v$ and $b_l$ for the double parabola model of
$\rho_{lv}(z)$ are obtained from a fit to the free liquid--vapor interface.
The film height for the model profile is then determined such that 
it matches the simulated profile exactly at $z=\ellav$, which, by construction, amounts to 
defining $\ellav$ such that the crossing criterion is met for both the model
and the simulated profiles. The predicted results are compared with simulations
in Fig.\ref{fig:profiles}, clearly showing  good agreement 
even for films as thin as $\ellav=1.9$ molecular diameters.

\subsection{Short range versus long range forces}

\label{sec:lvs}

In this chapter, we have seen how Density Functional Theory may provide accurate
and analytic results for density profiles of inhomogeneous fluids. It is also
pleasing to see that all such results, whether the shape of the liquid--vapor interface, the
density profile of a liquid in the neighborhood of a substrate, or the structure
of an adsorbed liquid film, are obtained within a consistent and unified
framework based on a single apparently general result, namely, \Eq{LDE}.

The density profiles for adsorbed films that are provided via \Eq{superpos}, may be
replaced back into the underlying density functional in order to obtain accurate
estimates of the film's free energy. Particularly, one can obtain from the free
energy functional the interface potential, i.e., the free energy of an adsorbed
film of height $\ellav$, measured relative to the free energy of an infinitely
thick film. To leading order, this is given by \cite{chernov88,henderson94}:
\begin{equation}\label{eq:gsr}
 g_{\rm sr}(\ellav) = -A_1 \cos(k_o\ellav+\theta_w) e^{-b_o\ellav} + A_2 e^{-2b_o\ellav}
\end{equation} 
where $A_1$ and $A_2$ are positive constants. Unfortunately, this model applies
for strictly short--range forces. This also includes the wall--fluid interactions,
which are considered in this expression as a contact potential of virtually zero range.

In practice, however, most fluids are subject to power--law interactions
that will decay as $r^{-6}$ or even slower  \cite{israelachvili91}. 
What is the status of our results then? Simple analytical expressions for
long--range fluids are extremely difficult to obtain already for van der
Waals interactions. Fortunately, one can still
work out their asymptotic decay \cite{barker82,dietrich91,mecke99b}. 
For example, the tails of the liquid--vapor
interface decay exponentially fast for the short--range fluids considered here, while they should
decay rather as $z^{-3}$ in a van der Waals fluid \cite{barker82}. 

Despite the omission of these fine details, the gross features of the interfacial structure 
are not expected to change significantly.
Indeed,  one can hardly expect that the neglect of power--law
tails of the fluid--fluid pair potential will upset the
packing effects that are observed at the wall--liquid interface; nor the fact that the 
interfacial width of the liquid--vapor interface decays in the scale of the correlation length.

These coarse structural details are many times  all what is needed to describe the most relevant
phenomenology. For example, in the thermodynamic perturbation theory of the
liquid--state, it suffices to provide the most crude approximation for the
structure of a hard sphere reference fluid in order to qualitatively account for
the role of dispersion forces. The rudimentary assessment of the first order
perturbation so achieved is sufficient to transform a dull monotonic equation of
state into a van der Waals isotherm exhibiting fluid coexistence
\cite{mcquarrie76}.

Similarly, the most significant feature of the van der Waals interactions of an
adsorbed fluid is the long range potential $V(z)=-\frac{\epsilon_w}{6\pi}z^{-3}$ that attracts the
fluid molecules towards the substrate \cite{israelachvili91}, producing an external field
contribution to the interface potential which is given by:
\begin{equation}\label{eq:gext}
  g_V(\ellav) = \int_{d_{w}}^{\infty} \left [ \rho_{\rm }(z;\ellav) - \rho_{\rm
  }(z;\ellav=\infty) \right
  ]V(z) d z
\end{equation} 
where $d_w$ is the distance of closest approach to the substrate.
Since, according to \Eq{generalbar}, the structure of the liquid film is hardly affected
by the external field, it suffices to consider  a simple step  like film profile in
order to assess the leading order contribution of the long--range interactions to
the interface potential: 
\begin{equation}
  \rho_{\rm }(z;\ellav) = \rho_l - \Delta\rho_{\rm lv}\, \mathcal{H}(z-\ellav) 
\end{equation} 
where $\Delta\rho_{\rm lv}=\rho_l-\rho_v$ and $\mathcal{H}(z)$ is the Heaviside function.
Substitution of this profile into \Eq{gext} readily yields the familiar Hamaker
long--range interaction of an adsorbed film:
\begin{equation}\label{eq:hamakerz2}
      g_V(\ellav) = \frac{H_w}{12\pi} \ellav^{-2}
\end{equation} 
where $H_w=\epsilon_w\,\Delta\rho_{\rm lv}$ is the Hamaker constant
\cite{israelachvili91}. Comparing the long--range decay of the above equation with the exponential
decay expected from \Eq{gsr}, one concludes that the signature of short--range
structural forces in the interface potential will be essentially washed--out by the van der Waals
interactions, as noted previously \cite{chernov88}.

\begin{figure}[t]
\centering
\includegraphics[clip,scale=0.30,trim=0mm 5mm 0mm 0mm,
angle=270]{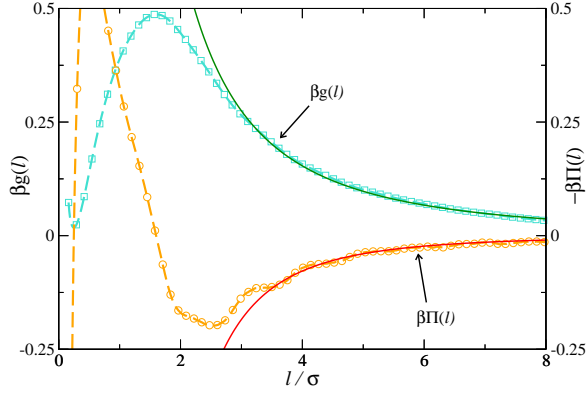}
\caption{Interface potential (squares) and disjoining pressure (circles) of a model of Argon
adsorbed on solid Carbon Dioxide. The dashed lines are a guide to the eye, while
the full monotonic lines depict the power law decay expected from the step
density model with van der Waals forces (results adapted from
Ref.\cite{macdowell13, gregorio12}, correspond to the same model and temperature as the three
preceding figures).
\label{fig:ifp_and_pi}
}
\end{figure}

Obviously, a more accurate expression is obtained if we use
\Eq{superpos} for the film profile. Unfortunately, the resulting integral does
not have a primitive that will provide us with much insight. It is more
convenient to consider a superposition approximation, but still using the
step--like model for the liquid--vapor interface, such that:
\begin{equation}
 \rho_{\rm }(z;\ellav) = [ 1 + h_{\rm wl}(z) ] [ \rho_l - \Delta\rho_{\rm
lv}
 \mathcal{H}(z-\ellav) ]
\end{equation} 
Rather than trying to solve for $g_V(\ellav)$, which is also quite
unpleasant, we consider the disjoining pressure, which may be obtained from
$g_V$ as:
\begin{equation}
   \Pi_V(\ellav) = -\frac{d g_V(\ellav)}{d \ellav} = - \int_{d_{\omega}}^{\infty} V(z)
   \frac{d \rho_{\rm }(z;\ellav)}{d \ellav}
\end{equation} 
whence, by virtue of the second equality, the Heaviside step function is
transformed into a Dirac function and projects the integrand out, yielding: 
\begin{equation}\label{eq:piosc}
   \Pi_V(\ellav) =   [ 1 + h_{\rm wl}(\ellav) ] \frac{H_w}{6\pi} \ellav^{-3}
\end{equation} 
It follows that, apart from  the well known leading order contribution to the
disjoining pressure of long--range fluids, our mean field calculations provide
an additional oscillatory contribution with a fast decay of order $e^{-b_o
\ell}/\ell^{3}$.
It is also worth noticing that, as successive derivatives of the interface
potential are performed, the decay of van der Waals tails becomes steeper,
whereas that of short range forces (c.f. \Eq{gsr}) remains of the same range.
Accordingly, it could be possible   to find a crossover from long--range to
short--range dominated interactions in either $\Pi(\ell)$ or its derivatives.
This  has been considered as a possible hypothesis for the
explanation of experimental findings that we will discuss later
\cite{werner99,sferrazza07}.

Figure \ref{fig:ifp_and_pi} shows computer simulation results for the
interface potential of Argon adsorbed on a solid substrate close to the wetting
temperature \cite{gregorio12}. The interface potential presents a minimum
corresponding to metastable equilibrium thin films, and a long--range monotonic decay
which, as shown in the figure, may be nicely described from the expected power
law of \Eq{hamakerz2}.
The disjoining pressure may be calculated from $g(\ell)$ by derivation
and is also shown in Fig.\ref{fig:ifp_and_pi}. Upon numerical derivation, the highly accurate data for $g(\ell)$
reveals  oscillatory behavior  completely washed out in the interface
potential. Such oscillatory behavior is the result of the layered structure of
the adsorbed films (c.f. Fig.\ref{fig:profiles}). The figure shows that the oscillations are
superimposed on the expected leading order monotonic decay of $\ell^{-3}$, as
suggested from \Eq{piosc}. 

The almost quantitative description of the density profiles afforded by \Eq{superpos},
and the qualitative description of the interface potential and disjoining
pressure afforded by \Eq{hamakerz2} and \Eq{piosc}, are a pleasing accomplishment of liquid state
theory.  If, however, we attempted to describe the oscillations exhibited by
the disjoining pressure from the known correlation function $h_{\rm wl}$, as
suggested by \Eq{piosc} we would find predicted oscillations with amplitudes
that are far too large. 

Why are the results of simulation so much smoothed relative to the theoretical
expectations of liquid state theory will be discussed in the next section.

\section{Classical Capillary Wave Theory}

In the previous section we have seen that Density Functional Theory  provides a
consistent and unified framework for the description of a purely flat interface,
where the density profile is only a function of the perpendicular direction, $z$.
Not unexpectedly, we have found that the structural properties of the interface,
as well as the interface potential and disjoining pressure are {\em intrinsic}
properties of the fluid, i.e.: they only depend on the fluid's structural
properties and on the intensive thermodynamic fields (temperature and chemical potential).

\begin{figure}[t]
\centering
\includegraphics[clip,scale=0.30,trim=0mm 5mm 0mm 0mm,
angle=0]{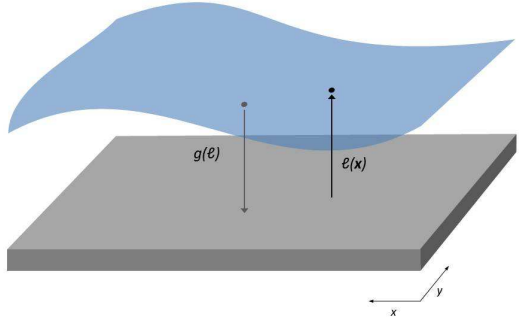}
\caption{Sketch of the rough interface of an adsorbed film subject to an
external field. For each point $\rpar$ on the plane of the
substrate, a film height $\ell(\rpar)$ is defined, and a local free
energy calculated as $g(\ell(\rpar)) d\rpar$.
\label{fig:monge}
}
\end{figure}

In practice, however, one can hardly expect the dividing surface of a film to
remain flat at finite temperature.
Rather, it is expected that thermal fluctuations will deform the interface, such 
that it becomes {\em rough} and accordingly the film profile deviates from its
average value. 
Such {\em capillary waves} may be described in terms of the Monge representation, where the film
thickness above a point $\rpar$ on a reference plane is given as a smooth
function $\ell(\rpar)$ (c.f. Fig.\ref{fig:monge}). Obviously, this description ignores overhangs and bubbles but
should be quite reliable away from the bulk critical point. Fluctuations of
$\ell(\rpar)$ away from the average increase the entropy of the interface, but are at the
cost of increasing the surface area. 
Furthermore, whether the film is adsorbed on a substrate, or subject to the
effect of gravity, it will feel an external field that restricts the fluctuations of
$\ell(\rpar)$ via the interface potential $g(\ell)$. This physical situation
clearly calls for a description in terms of the Interfacial Hamiltonian
described in the introduction, \Eq{ifh}. In this context, we will also refer to
IHM, as the capillary wave Hamiltonian, CWH.

In order to avoid confusion between the rough interface profile, $\ell(\rpar)$,
and its thermal average, denoted  $\ell$ in the previous section, we will
usually refer to  $\ell(\rpar)$ as $\Sigma$ (for sinusoidal interface).
In some instances, we will also use the label $\pi$ (short for planar) in order
to stress the difference between properties of a rough interface and those of an
assumed planar interface.

\subsection{Capillary wave spectrum}

\label{sec:ccws}

Let us now consider to what extent do thermal capillary waves modify the
structure of the interface as described in the previous section. 
As we will show, the consequences are actually very
important, already at the lowest order of approximation.

First, we consider the limit
of small gradients,  $(\nabla \ell)^2\ll 1$. This allows us to get rid of the
unpleasant square root, and write:
\begin{equation}\label{eq:cwhlin}
H[\Sigma] = \int d \rpar \left \{ g(\ell(\rpar)) + \frac{1}{2}\gamma_{\infty}  [
\nabla \ell(\rpar) ]^2    \right \}
\end{equation} 
where, as explained above, $\Sigma$ is a shorthand for the functional dependence
of $\ell(\rpar)$.
Despite the apparently very different physics, the capillary wave Hamiltonian is
formally identical to the square gradient functional (c.f. \Eq{Fsgt}), with
$\ell(\rpar)$ playing the role of
$\rho(\rvec{}{})$ and $\gamma_{\infty}$ the role of $C_{\infty}$. Relating $\rho(\rvec{}{})$
with $\ell(\rpar)$ allows us to transform the gradient of densities into a gradient
of $\ell(\rpar)$ and identify the Square Gradient Functional with 
$H[\Sigma]$ \cite{jin93,safran94} to leading order, as we shall see in section \ref{sec:ncwt}.

In order to proceed, we expand the integrand in small deviations away from the
average film height. Defining $\delta\ell(\rpar)= \ell(\rpar) - \ellav$, and performing a Taylor
series of $g(\ell)$ leads to:
\begin{equation}\label{eq:pr1}
H[\Sigma] = \int d \rpar \left ( g(\ellav) + g'(\ellav) \delta \ell + \frac{1}{2} g''(\ellav) \delta\ell^2 
   + \frac{1}{2}\gamma_{\infty}  ( \nabla \ell(\rpar) )^2    \right )
\end{equation} 
It is now convenient to describe the film height fluctuations in terms of  Fourier modes,
$\delta\ell_{\qvec}$, as follows:
\begin{equation}\label{eq:lfourier}
  \delta\ell(\rpar) = \sum_{\qvec} \delta\ell_{\qvec} e^{i\qvec\cdot\rpar}
\end{equation} 
Plugging this result back into equation \Eq{pr1}, followed by some rearrangements, then
yields:
\begin{equation}\label{eq:pr2}
\begin{array}{lll}
H[\Sigma] & = & A g(\ellav) +  \int d \rpar \left [ g'(\ellav) \sum_{\qvec}
\delta\ell_{\qvec}\, e^{i\qvec\cdot\rpar} + \right . \\
 & & \\
 & & 
\left . \frac{1}{2} \sum_{\qvec} \sum_{\qvec'} \left ( g''(\ellav)  
   + \gamma_{\infty}\, \qvec\cdot\qvec' \right ) \delta\ell_{\qvec}
   \delta\ell_{\qvec'}\,   e^{i(\qvec+\qvec')\cdot\rpar}    \right ]
\end{array}
\end{equation} 
where $A$ is the surface area of the flat interface.
The integral over $\exp(i\qvec\cdot\rpar)$ is $A\,\delta_{\qvec, 0}$,
while that over
$\exp{\left [i(\qvec+\qvec')\cdot\rpar \right ]}$ is likewise
$A\,\delta_{\qvec,-\qvec'}$, with $\delta_{\qvec,\qvec'}$,
Kronecker's delta 
\cite{goldenfeld92}.  Furthermore, we take into account that by definition, $\delta
\ell(\rpar)$ describes fluctuations about the average film height, so that the zero
wave vector mode  $\delta\ell_{\qvec=0}$ is null.
With this in mind, we can now integrate \Eq{pr2}, to obtain:
\begin{equation}\label{eq:cwhfour}
  H[\Sigma] = A\left \{ g(\ellav) + \frac{1}{2} \sum_{\qvec} [ g''(\ellav) + 
   \gamma_{\infty} q^2 ]
 | \delta\ell_{\qvec}  |^2 \right \}
\end{equation} 
This result provides us with the free energy of a frozen realization of the
interfacial roughness. We can define the probability of such realization with the
usual Boltzmann weight:
\begin{equation}
   P(\Sigma) = \frac{ e^{-\beta H (\Sigma ) } }{ Z_{cw} }
\end{equation} 
where $Z_{cw}$, the partition function, is now a sum over all possible capillary wave
realizations. In terms of the capillary wave modes, this can be written as:
\begin{equation}
  Z_{cw} = \int \prod d \ell_{\qvec}\, e^{-\beta H[\Sigma] }
\end{equation} 
Since $H$ is given in terms of independent additive Fourier mode contributions, it
can be factored into a product of simple integrals as in the case of the partition
function of an ideal gas, so that we can write:
\begin{equation}
    Z_{cw} = e^{-\beta A g(\ellav)} \prod_{\qvec} \int d \ell_{\qvec} 
				e^{-\frac{1}{2}\beta A  [ g''(\ellav) + \gamma_{\infty} q^2 ] |\delta\ell_{\qvec}|^2 }
\end{equation} 
Taking into account that $|\delta\ell_{\qvec}|^2$ is actually the
complex squared modulus of $\delta\ell_{\qvec}$, it follows that the integral is of
Gaussian form. Despite some subtleties related to integration in the complex plane
\cite{kayser86,goldenfeld92}, it satisfies the equipartition theorem. Considering
$|\delta\ell_{\qvec}|^2$
to play the role of squared velocity and $A [ g''(\ellav) + \gamma_{\infty} q^2 ]$ the role
of mass, we can then write:
\begin{equation}\label{eq:ccws}
      A \langle \delta\ell_{\qvec} \delta\ell_{-\qvec} \rangle = \frac{k_BT}{
	  g''(\ellav) + \gamma_{\infty} q^2  }
\end{equation} 
where the angle brackets denote a thermal average. 

\begin{figure}[t]
\centering
\includegraphics[clip,scale=0.30,trim=0mm 5mm 0mm 0mm,angle=270]{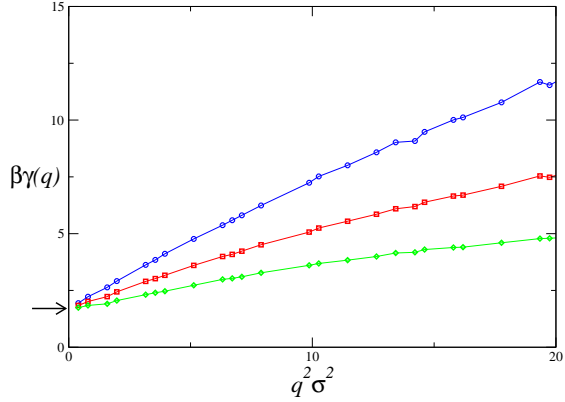}
\caption{Plot of the effective wave--vector dependent surface tension (left hand
side of \Eq{gqf}). Results are displayed for the liquid--vapor interface of a
Lennard--Jones model of Argon close to the triple point (the model and
temperature are as studied in Ref.\cite{macdowell13}, employed in Fig.2--5 and described in
more detail in sec.\ref{sec:simnew}). The symbols are simulation
data obtained for three different choices of the interface position, and the
lines are a guide to the eye. The arrow indicates independent results
for the liquid--vapor surface tension \cite{gregorio12}.
\label{fig:gammaq2}
}
\end{figure}

This result states that the
mean squared amplitude of the Fourier modes  decreases as the square of the
in--plane wave vector increases. Such expectation has been confirmed in a great number of
computer simulation studies for the special cases of {\em free--interfaces}, 
i.e., in the absence of external fields, whence $g''(\ellav)=0$.
In this simple case, one can arrange \Eq{ccws} as:
\begin{equation}\label{eq:gqf}
  \frac{1}{A q^2 \langle|\delta\ell_{\qvec}|^2\rangle} = \beta\gamma_{\infty}  
\end{equation} 
It follows that a plot of the left hand side as a function of
$q^2$ is a constant equal to the surface tension. In practice, for the
small systems that are usually considered in computer simulation studies, it is
difficult to achieve the regime where \Eq{gqf} is actually constant, but the
results may be safely extrapolated to $q=0$ and provide good estimates of the
surface tension \cite{mueller96,mueller96b,mueller00,milchev02} or even the
stiffness of solid--fluid interfaces above the roughening transition
\cite{davidchack06,zykova-tilman10,rozas11}.

For the large $q$ regime that is achievable in
simulations, it is found that the left hand side of \Eq{gqf} provides a
phenomenological definition for a wave vector dependent surface tension
$\gamma(q)$, describing the deviations of $\langle|\delta\ell_{\qvec}|^2\rangle^{-1}$ from
the expected low $q$ regime of $\beta\gamma_{\infty}$ \cite{mecke99b,mora03}. From theoretical
considerations, it is known that the linear term in $\gamma(q)$ is absent, so
that, to lowest order, one can write \cite{mecke99b,mecke01}:
\begin{equation}\label{eq:gammaq}
 \gamma(q) = \gamma_{\infty} + \kappa q^2
\end{equation} 
where $\kappa$ is known as the {\em bending rigidity}.

Figure \ref{fig:gammaq2} shows a plot of the left hand side of \Eq{gqf}, as a
function of $q^2$, as obtained for a liquid--vapor interface of a Lennard--Jones
model of Argon. Since the definition of the interface down to atomic length
scales has some degree of arbitrariness (this will be discussed at length in
section \ref{sec:location}), the spectrum depends on the actual criteria that are employed
to locate it. Results for three different choices of the interface position are
presented in the figure. Whereas all such choices yield different results, it
is clear that 1) all the spectra extrapolate to the same $q=0$ value, coincident
with the liquid--vapor surface tension obtained independently and indicated in
the figure with an arrow and 2),
in a regime of long wave--vectors up to about $q=2$ inverse molecular diameters,
a clearly linear behavior is found consistent with \Eq{gammaq}.

Two important considerations are worth mentioning at this stage:

1) Exactly how the square gradient coefficient of the CWH, $\gamma_{\infty}$ is
related to the actual liquid--vapor surface tension has been the matter of
debate for a long time (see an interesting review by Gelfand and Fisher for
further details on this issue \cite{gelfand90}). In the original formulation of Buff, Stillinger and
Lovett \cite{buff65}, $\gamma_{\infty}$ was considered an effective free energy,
smaller than the
experimental tension by terms of order $q_{\rm max}^2$, with $q_{\rm max}$ an
upper cutoff wave--vector. However,
ample theoretical and simulation evidence has gathered favoring the
interpretation of $\gamma_{\infty}$ as the actual experimentally accessible liquid--vapor surface tension
\cite{weeks77,abraham81,mueller00,vink05},
or at least, as a finite system size approximation
\cite{kayser86,jasnow84,gelfand90}. The explicit form of the
finite size dependence is also a matter of debate, but simulation results
suggest that the dependence is weak \cite{binder82,chen95,aguado01b}. For this reason, in what follows we will
refer to $\gamma_{\infty}$ as the macroscopic liquid--vapor surface
tension. Rather, anticipating results of section \ref{sec:ncwt}, we
employ the subindex $\infty$, in order to stress  we refer here to the
surface tension in the absence of an external field, i.e. at infinite distance
away from the field.

2) Intuitively,  a positive slope of the phenomenological $\gamma(q)$ is in principle expected,
since, a negative slope would imply apparently unphysical diverging low wavelength
interfacial fluctuations. However, it has been suggested that fluids with
van der Waals interactions have a negative effective bending rigidity, 
with $\gamma(q)$ exhibiting a minimum at finite $q$ and then increasing as
expected for large $q$ \cite{mecke99b}. This hypothesis, which has been supported
by experiments \cite{mora03}, is however still to date subject to some
reservations \cite{tarazona07,paulus08,pershan12}. Clearly, the results of
Fig.\ref{fig:gammaq2} illustrate the difficulties of defining unambiguously a
bending rigidity, since it depends on the
somewhat arbitrary procedure employed for the precise location of the interface
\cite{tarazona07,blokhuis09}.

\subsection{The interfacial roughness}

\label{sec:roughness}

Unfortunately, except for very few instances \cite{mora03} and some reservations
\cite{paulus08}, grazing x--ray
scattering studies
do not provide the resolution that is required to test the full capillary wave
spectrum, i.e., \Eq{ccws} (c.f. Ref.\cite{pershan12} and \cite{daillant00} for
reviews on x--ray scattering studies of surfaces). Rather, it is the interfacial roughness 
$\Delta_{\rm cw}^2 = \langle\delta \ell^2\rangle$ that is usually measured,
whether one considers the free fluid interface \cite{schwartz90,sanyal91,ocko94}, 
or that of adsorbed films \cite{tidswell91b,doerr99,heilmann01,plech02}.

Using Plancherel's theorem, it is possible to relate the  lateral average of
$\delta \ell^2$ with that of $\delta\ell_{\qvec}\cdot \delta\ell_{-\qvec}$, so
that the  roughness may be  determined by summation
of the thermally averaged squared Fourier modes as:
\begin{equation}
  \Delta_{\rm cw}^2 = \sum_{\qvec} \langle \delta\ell_{\qvec}\cdot
\delta\ell_{-\qvec}\rangle 
\end{equation} 
Considering the transformation $\sum_{\qvec} \rightarrow \frac{A}{4\pi^2}
\int\int d q_x d q_y$
we can evaluate the interfacial roughness as the integral of \Eq{ccws}:
\begin{equation}\label{eq:intrough}
  \Delta_{\rm cw}^2 = \frac{k_B T}{2\pi} \int_{q_{\rm min}}^{q_{\rm max}} 
  \frac{q\, d q}{ g''(\ellav) + \gamma_{\infty} q^2 }
\end{equation} 
where, by virtue of the isotropy of the interface in the transverse direction,
we have transformed $d q_x d q_y$ into $2\pi q d q$. The lower bound of the
integral $q_{\rm min}=2\pi/L$ is given by the finite system size of the
simulation, or by the experimental setup. Unfortunately, the integral does not
converge, and an {\em ad hoc} maximum wave--vector cutoff $q_{\rm max}$ has to be
introduced. This is not always a problem in experimental studies, since  the
maximal wave--vector can be identified with an instrumental cutoff related to
the maximal momentum transfer, but does become
an unpleasant problem in simulation studies, where the resolution goes down to the
atomic scale. In practice, one assumes $q_{\rm max}=2\pi/\lambda_{\rm min}$,
with $\lambda_{\rm min}$ an empirical parameter which has been interpreted
either as an atomic length scale \cite{ocko94}, or the bulk correlation length
\cite{kayser86,sengers89}. The difference
is of little consequence at low temperature, but should be a matter of
concern as the critical point is approached.

Performing the integral, \Eq{intrough}, we obtain  finally the capillary--wave--induced interfacial roughness:
\begin{equation}\label{eq:ccwr}
 \Delta^2_{cw} = \frac{k_B T}{4\pi\gamma_{\infty}} 
   \ln \frac{1 + \xi_{\parallel}^2 q_{\rm max}^2}{1 + \xi_{\parallel}^2 q_{\rm min}^2}
\end{equation} 
where the relevant length scale here:
\begin{equation}
\xi_{\parallel} = \sqrt{\frac{\gamma_{\infty}}{g''}}
\end{equation} 
is known as the parallel correlation length and dictates the range of capillary wave fluctuations in the
transverse direction \cite{henderson92,evans92}. For liquid--vapor interfaces
under gravity, $\xi_{\parallel}$ may be immediately identified with the
capillary length, $a$. For films adsorbed on a substrate,  it is also sometimes
known as the {\em healing distance} \cite{degennes04}, and dictates the ability of a liquid film
to match the roughness of the underlying substrate
\cite{robbins91,tidswell91,degennes04}. On the other hand, $\Delta_{\rm cw}^2$
is also some times known as the perpendicular correlation length, and written
alternatively as $\xi_{\perp}$. Table \ref{interaction_list} provides a
list of parallel and perpendicular correlation lengths for different important intermolecular
forces acting on the liquid--vapor interface.

\begin{table}
\begin{center}
\begin{tabular}{cc|cccc}
\hline
\toprule
 & & \multicolumn{4}{c}{External Field} \\
\cmidrule{3-6}
 & & System Size & Short Range & van der Waals &  Gravity \\
\cmidrule(r){1-2}  \cmidrule(r){3-3}  \cmidrule(r){4-4}
\cmidrule(r){5-5} \cmidrule(r){6-6}
  & & & &  & \\
 $g(\ell)$ & & - & $D\exp(-b\ell)$ & $\frac{H_w}{12\pi}\ell^{-2}$ &  $\frac{1}{2}m\Delta\rho G \ell^2$ \\
  & & & & \\
 $\xi_{\parallel}^2$ & & $\gg L^2$  & $\frac{\gamma_{\infty}}{D b^2}\exp(b \ell)$ &
 $\frac{2\pi\gamma_{\infty}}{H_w}\ell^4$ & $\frac{\gamma_{\infty}}{m\Delta\rho G}$ \\
  & & & &  & \\
$\Delta_{\rm cw}^2$ & & $\ln L$ & $\ell$ & $\ln \ell$ & $\ln a$ \\
\bottomrule
\end{tabular}
\end{center}
\caption{\label{interaction_list} Table of interface potentials, $g(\ell)$ and parallel correlation
lengths, $\xi_{\parallel}$ for common external fields acting on an interface. The last row indicates the
leading order dependence of the capillary roughness, $\Delta_{\rm cw}^2$ that results. $D$ is the amplitude of
short range forces, and has dimensions of energy per unit area. When $1/b$ is
identified with the Debye screening length, the results may describe interactions
arising from the electric double layer.$H_w$ is the Hamaker constant. 
$a^2=\gamma_{\infty}/m\Delta\rho G$, with $G$ the acceleration of
gravity, is the squared capillary length.}
\end{table}

The above result is of experimental relevance in two limiting cases. 

\paragraph{Weak fields}
For very weak external fields, the parallel correlation length becomes very
large, and may actually achieve values much larger than
the  lateral system size. In this case,  $\xi_{\rm \parallel}\, q_{\rm min} \gg 1$ 
and  the capillary wave roughness becomes:
\begin{equation}\label{eq:wfrough}
   \Delta^2_{cw} = \frac{k_B T}{2\pi\gamma_{\infty}}
   \ln \frac{q_{\rm max}}{q_{\rm min}}
\end{equation} 
This result implies a logarithmic dependence on system size $L$ (simulations) or
experimental lower cutoff $\lambda_{\rm min}$ which has been fully confirmed.
The most natural way to study this limit is a computer simulation study, where
one can prepare a liquid slab inside a simulation cell at zero field. In
practice, however, the capillary length for essentially all liquids is so  much
larger than the upper wavelength cutoff  afforded with scattering techniques that 
also ordinary fluid interfaces under the effect of gravity are in this limit.
Indeed, both computer simulations  \cite{lacasse98,sides99,vink05,geysermans10} 
and experimental studies \cite{schwartz90,carelli05,doerr99} agree as to
the logarithmic dependence of the interfacial roughness, and confirm that the
slope of $\Delta^2_{cw}$ as a function of $\ln q_{\rm min}$ yields a reliable
estimate of the surface tension
\cite{lacasse98,sides99,vink05,luo06,werner97,geysermans10,rozas11}.

\paragraph{Strong fields} 
If, on the other hand, the interface is subject to a strong field, as is the
case for a thin adsorbed film subject to a disjoining pressure, the parallel
correlation length is small but $\xi_{\rm \parallel} q_{\rm max}$ usually remains much
larger than unity. In most practical realizations, however $\xi_{\rm \parallel}
q_{\rm min} \ll 1$ 
and, as a result the roughness is no longer system size dependent:
\begin{equation}\label{eq:ccwrsf}
  \Delta^2_{cw} = \frac{k_B T}{2\pi\gamma_{\infty}}
          \ln (\xi_{\parallel} q_{\rm max} )
\end{equation} 
In this equation, $\xi_{\parallel}$ plays a similar role as $q_{\rm min}^{-1}$ in
the weak field limit. Also in this case, there is a large amount of evidence
strongly in favor of a logarithmic dependence of $\Delta^2_{cw}$ on
$\xi_{\parallel}$. In most practical realization, the adsorbed liquid film is
subject to van der Waals forces, so that the liquid--vapor interface is bound by
an interface potential $g(\ell)\propto \ell^{-2}$. As a result, the interfacial
roughness exhibits a logarithmic dependence on the film thickness (c.f. Table \ref{interaction_list}),
and a fit to
the experimental data actually provides reasonable estimates of the Hamaker
constant \cite{tidswell91,doerr99,sferrazza97,heilmann01,plech02}. An even more
striking confirmation of this result is afforded in systems where the Hamaker
constant is very small. In such cases, the dominant contribution stems from
short range forces. The interface potential is now of the form, $g(\ell)\propto e^{-\ell}$, 
so that the capillary wave roughness  grows as the square root of the film
thickness increases,
as illustrated in Table \ref{interaction_list} \cite{parry92,kerle96,werner97}. 

Despite this amount of experimental evidence, the situation of \Eq{ccwr} seems far
less satisfactory in the strong field limit than it is for the weak field limit.
Indeed, many studies report a capillary roughness that is either too large
\cite{plech02} or too small \cite{heilmann01} relative to expectations from
\Eq{ccwrsf},
while other studies find the logarithmic prefactor incompatible with the known
interfacial tension \cite{heilmann01,doerr99}. In some instances, these discrepancies 
have been attributed to a possible cross--over from long--range
($\xi_{\parallel}\propto \ell^{-4}$) to short--range forces
($\xi_{\parallel}\propto \exp(-\ell)$) \cite{werner99,sferrazza07}; while in others it
has been suggested the need to somehow incorporate a film--thick--dependent
interfacial tension \cite{werner99,heilmann01}. Be as it may, the long
wavelength dependence of \Eq{ccwr} is
essentially uncontested and remains to date the framework for experimental
analysis.

As a final comment, it is worth mentioning that considering explicitly the
wave--vector dependent surface tension as dictated by \Eq{gammaq} into the capillary
spectrum of \Eq{ccws}, would in principle allow to eliminate the need for an
empirical upper wave--vector cutoff. Indeed, the Fourier amplitudes are then
given by:
\begin{equation}\label{eq:ccwskappa}
      A \langle \delta\ell_{\qvec} \delta\ell_{-\qvec} \rangle = \frac{k_BT}{
	  g''(\ellav) + \gamma_{\infty} q^2 + \kappa q^4  }
\end{equation} 
For positive $\kappa$ at least, the integral now converges and needs not the
upper cutoff \cite{daillant00,mecke01,pershan12}. Unfortunately, the resulting expression, which
is far less convenient, has been seldom employed \cite{daillant00,luo06b,hou13}.

\subsection{Intrinsic and capillary wave broadened profiles}

The prediction of large perpendicular interfacial fluctuations $\Delta^2_{\rm cw}$ for
a liquid--vapor interface poses a serious challenge to the traditional view of a
well defined, {\em intrinsic density profile}, say, $\rho_{\pi}(z)$, as
described in section \ref{sec:dft}.  According to the
picture that emerges from \Eq{ccwr}, the liquid--vapor interface of a substance on
earth exhibits almost unbound perpendicular fluctuations up to the capillary
length, which, for a fluid such as water at ambient temperature is on the $mm$ length scale. This
implies that  a fixed point $z$ say a $\mu m$ away from the equimolar dividing surface,
is found alternatively within the liquid or vapor phases, such that its average
density is simply half way between $\rho_l$ and $\rho_v$. A pessimistic
interpretation of this result is that, in the absence of the gravitational field
the liquid-vapor interface cannot possibly exist in the thermodynamic limit.
This view relies too heavily on the significance of averages, particularly those
collected over an infinite period of time (an analogous interpretation would imply that
Brownian particles do not move, because their average position is zero, whereas
we know that the most likely event is that each such particle in a sample would have moved
away from the original position as $\sqrt{t}$). In practice, the relaxation dynamics of the
capillary waves is also very slow \cite{jeng98} and there is no problem in
identifying the interface over long but finite periods. Furthermore, it has been
shown that the intrinsic
density profile is always recognizable at the scale of the bulk correlation
length, provided it is measured relative to the instantaneous interface
position \cite{weeks77}.  

\begin{figure}[t]
\centering
\includegraphics[clip,scale=0.30,trim=0mm 5mm 0mm 0mm,
angle=270]{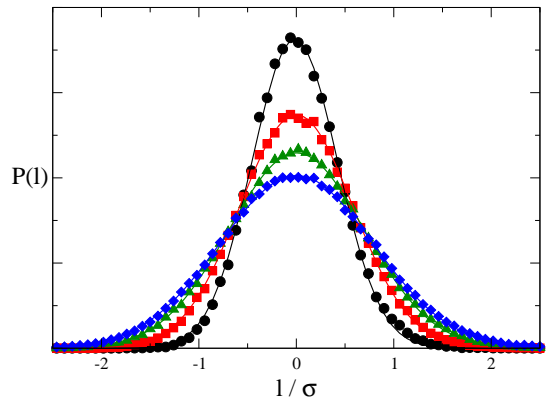}
\caption{Probability distribution of local film thickness of a liquid--vapor
interface at moderate temperature as described by the Lennard--Jones 
model of Argon. 
Results are depicted for system sizes with lateral area of $100$ (black), $200$
 (red), $300$ (green) and $400$ (blue)   squared molecular diameters, $\sigma$. Symbols are results from
simulation, while the lines are Gaussian fits (results from Ref.\cite{benet13}
at $k_BT=0.90\epsilon$, c.f. Sec.~\ref{sec:simnew} for more details on the model).
\label{fig:pl}
}
\end{figure}

For small systems, however, even a small observation time as is usually afforded
in computer simulations or x-ray scattering experiments   produces an averaged
density profile which exhibits the fingerprints of capillary wave broadening.
The connection between the average density profile that can be actually measured and
the underlying intrinsic density profile relevant to length--scales below the
bulk correlation length may be performed by means of a convolution
\cite{weeks77,bedeaux85b,jasnow84,benjamin92}, as we shall soon see. First, however,
consider the picture that emerges from the capillary wave theory:
a local displacement of the interface about its average, $\delta\ell(\rpar)$, translates the
whole density profile by exactly that amount, such that
the instantaneous density  becomes
$\rho_{\pi}(z-\delta\ell)$. To see how this changes the measured average density,
let us Taylor expand the translated profile about $z$:
\begin{equation}
  \rho(\rvec{}{};\Sigma) = \rho_{\pi}(z) - \rho'_{\pi}(z) \delta\ell(\rpar) +
  \frac{1}{2} \rho''_{\pi}(z) \delta\ell(\rpar)^2 
\end{equation} 
Performing now a lateral average, the linear term in $\delta\ell(\rpar)$ vanishes, but
the quadratic term does not. It is then apparent   that, for a fluctuating
interface, the averaged density $\rho(z;\Sigma)$ cannot possibly be equal to
the intrinsic density profile, but rather is:
\begin{equation}\label{eq:rhobexp}
  \rho(z;\Sigma) = \rho_{\pi}(z) + \frac{1}{2} \rho''_{\pi}(z)
\langle\delta\ell(\rpar)^2 \rangle
\end{equation} 
Since, for reasons of symmetry, the probability of exhibiting translations to
the left or to the right must be equal for a free interface, we expect a
Gaussian distribution for $\ell(\rpar)$,  with width equal to $\langle \ell(\rpar)^2
\rangle$. Whence, alternatively to the series representation of \Eq{rhobexp}, we can write
the effect of the interface translations by means of  a convolution, as follows
\cite{weeks77,bedeaux85b,jasnow84,benjamin92}:
\begin{equation}\label{eq:cconv}
    \rho(z) =   \int \rho_{\pi}(z-\ell) P(\ell) d \ell
\end{equation} 
where $P(\ell)$ is the probability density for the interface displacements. The
theoretical expectation of a Gaussian distribution of  width $\Delta_{\rm cw}$
has been convincingly confirmed in numerous computer simulation studies
\cite{benjamin92,werner97,vink05}, so that we can safely assume
\cite{weeks77,bedeaux85b}:
\begin{equation}\label{eq:cgauss}
   P(\delta\ell) = \frac{1}{\sqrt{2\pi\Delta_{\rm cw}^2}} e^{-\frac{1}{2} \frac{
   \delta\ell^2 }{ \Delta_{\rm cw}^2 } }
\end{equation} 
Fig.\ref{fig:pl} displays the probability distribution of the local interface
position for the liquid--vapor interface of a Lennard--Jones like Argon model at
moderate temperature. The results are given for systems with increasing lateral
system size and
clearly show that the probability distribution is Gaussian, becoming
flatter as the lateral area increases.

The role of capillary roughening is perhaps best illustrated using the most crude
possible description for the intrinsic profile, i.e., a simple step function of
the form:
\begin{equation}
\rho_{\pi}(z) = \frac{1}{2}(\rho_l + \rho_v) -  \frac{1}{2}(\rho_l - \rho_v) (2
\mathcal{H}(z)-1)
\end{equation}  
with $\mathcal{H}(z)$ the Heaviside function. The convolution of \Eq{cgauss} transforms the
discontinuous step--like density profile into a smooth error function 
of width $\sqrt{2}\,\Delta_{\rm cw}$ \cite{bedeaux85b}:
\begin{equation}\label{eq:cerf}
 \rho(z) = \frac{1}{2}(\rho_l + \rho_v) -  \frac{1}{2}(\rho_l - \rho_v)
 {\rm Erf}(\frac{z}{\sqrt{2}\,\Delta_{\rm cw}})
\end{equation} 
It is a remarkable
achievement of mathematical physics to show that, for the free interface of
a two dimensional Ising model, \Eq{cerf}, with $\Delta_{\rm cw}$ given by
\Eq{wfrough}, follows exactly from the underlying
microscopic Hamiltonian \cite{abraham81,fisher82}.

\begin{figure}[t]
\centering
\includegraphics[clip,scale=0.30,trim=0mm 0mm 0mm 0mm,
angle=270]{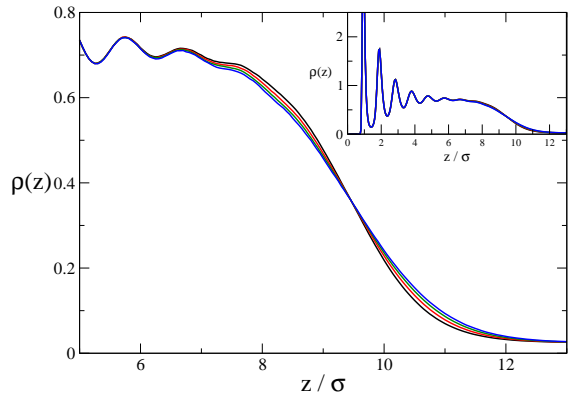}
\caption{
Detailed view of the liquid--vapor density profile of an adsorbed Lennard--Jones fluid above the wetting
transition.  Results are given for systems with lateral area increasing from 100
to 400 squared molecular diameters, $\sigma$, in the order black, red, green, blue. The
inset displays the overall view of the density profile, where it becomes clear
that the broadening only affects the liquid--vapor interface, but not the
layered structure (results from \cite{benet13}, the model is as that employed in
\cite{macdowell13} and described in Sec.~\ref{sec:simnew}, with an increased wall strength of
$\epsilon_w=143\epsilon\sigma^3$ and a temperature of $k_BT=0.90\epsilon$).
\label{fig:rhorough}
}
\end{figure}

According to the above equation, in the weak field limit, where $\Delta^2_{\rm cw}\propto \ln L$, the averaged
profile becomes completely smoothed out for $L \to \infty$. 
For finite system sizes, the effect is also apparent and measurable. Figure
\ref{fig:rhorough} displays density profiles for adsorbed films of the
Lennard--Jones Argon model above the wetting transition for several system sizes. 
The film is stratified as is usual for atomic fluids (inset), but a closer look
clearly shows how the liquid--vapor interface decays over larger and larger
length scales as the lateral system size is increased. The broadening of the
density profile may be measured and tested against expectations from \Eq{ccwr},
providing an independent means of estimating the surface tension
\cite{lacasse98,sides99,vink05,luo06,mitrinovic00,werner97,geysermans10,rozas11}.

Actually, the result of \Eq{cerf} serves as starting point for the analysis of most experimental
studies on capillary waves
\cite{pershan12,schwartz90,sanyal91,tidswell91b,ocko94,sferrazza97,doerr99,daillant00,heilmann01,plech02,carelli05,luo06b,sferrazza07}.
Low grazing beams on a surface  produce scattering intensities which probe the
density profile along a direction perpendicular to the interface. 
For the simple case of a single moderately rough interface, the reflectivity is given 
by \cite{pershan12,tidswell91b,heilmann01}:
\begin{equation}
 R(Q) \propto \left | \int d z \frac{d \langle\rho\rangle}{d z} e^{i Q z} \right |^2
\end{equation} 
where $Q$ is a scattering vector. Using \Eq{cerf} into the scattering formula, it
is found that the capillary waves result in a Debye--Waller like attenuation
factor for the reflectivity:
\cite{pershan12,ocko94,sferrazza97,luo06}:
\begin{equation}
 R(Q) \propto e^{-Q^2 \Delta_{\rm cw}^2 } 
\end{equation} 
Accordingly, a plot of $\ln R(Q)$ against $Q^2$ provides a straight line with a
slope equal to the capillary roughness. 
In practice, however, the intrinsic density profile is not an infinitely steep
step function, but has its own {\em intrinsic width}. As a result, the
interfacial width that is measured in scattering experiments has both intrinsic
and capillary wave contributions, and one usually assumes that the actual
measured roughness, say, $\Delta_{\rm exp}$ is given by
\cite{ocko94,sferrazza97,heilmann01,plech02,luo06b}:
\begin{equation}
 \Delta_{\rm exp}^2 = \Delta_{\pi}^2 + \Delta_{\rm cw}^2
\end{equation} 
where $\Delta_{\pi}^2$ is attributed to the intrinsic density profile. This is not
an all together convenient situation, since $\Delta_{\pi}^2$ cannot be measured
independently, so that it adds to the upper cutoff $q_{\rm max}$  yet another
empirical parameter.  Unfortunately, one cannot
actually resolve $\Delta_{\pi}$ from the $\ln q_{\rm max}$ contribution of
$\Delta_{\rm cw}$, and there are no other ways to distinguish from one another
than plausible arguments. In principle, the situation could be remedied by assuming a
reasonable intrinsic density profile, and performing the convolution of
\Eq{cconv}.
In practice, however, the convolution cannot be obtained analytically, not even
for the simple $\tanh(z)$ function, and only
occasionally it is performed numerically
\cite{werner97,mueller00,carelli05,luo06}. For that reason,
either ${\rm Erf}$ or $\tanh$ functions are employed to
fit the density broadened profiles, and $\Delta_{\rm cw}$ is obtained as a
fitting parameter.  There is however some evidence that the ${\rm Erf}$ profile
is a better choice \cite{sides99}.

Surprisingly, the fact that the double--parabola model provides an analytic
expression for the convolution has not been recognized. Indeed, plugging
\Eq{dplv}
into \Eq{cconv}, and using \Eq{cgauss}, yields, for the broadened density
profile the following lengthy but convenient result:
\begin{equation}
 \rho(z) = \rho_{\rm step}(z) + \Delta\rho_{\rm str}(z)
\end{equation} 
where $\rho_{\rm step}(z)$ is the leading order broadening from the
structure--less step model
(\Eq{cerf}), and $\Delta\rho_{\rm str}(z)$ is the additional contribution due to the
intrinsic interfacial structure \cite{palanco13}:
\begin{equation}
\begin{array}{lll}
 \Delta \rho_{\rm str}(z) & = & 
 \frac{1}{2} \frac{\rho_l - \rho_v}{b_l + b_v} 
 \left \{ 
   b_l e^{-b_v(z-\frac{1}{2} b_v \Delta_{\rm cw}^2)}
 {\rm Erfc}\left ( -\frac{z - b_v\Delta^2_{\rm cw}
 }{\sqrt{2}\Delta_{\rm cw}} \right ) \right . \\
  & & \\
 & & -
  \left . b_v e^{b_l(z+\frac{1}{2} b_l \Delta_{\rm cw}^2)}
 {\rm Erfc}\left (  \frac{z + b_l\Delta^2_{\rm
 cw}}{\sqrt{2}\Delta_{\rm cw}} \right )
 \right \}
\end{array}
\end{equation} 
Using this equation to fit the reflectivity data would allow to resolve the
capillary wave roughness $\Delta_{\rm cw}$ from the intrinsic structure, as well
as to obtain an estimate of the bulk correlation lengths. One could then extract
$q_{\rm max}$ meaningfully from  $\Delta_{\rm cw}$, and compare
$\lambda_{\rm min}=2\pi/q_{\rm max}$  with the bulk correlation
lengths obtained from the fit.

To sum--up, we have shown that thermal capillary waves considerably modify the
structure of the interface, conveying information on the whole system size into
the otherwise {\em intrinsic} density profile. We have shown that
the predictions of the classical capillary wave theory seem very well tested for
interfaces subject to a weak field, but some discrepancies seem to arise for
adsorbed films subject to relatively strong fields. Unfortunately, comparisons
between theory and experiment are not straightforward, because experimentally
only the capillary roughness is usually accessible. A much more stringent test
could be achieved if the whole capillary wave spectrum of adsorbed films were
measured. Since x--ray measurements are still very difficult to achieve at the
level of resolution that is sought, computer simulations would seem an ideal
tool. In the next section we will review current state of the art methods for
the computer simulation of the spectrum of adsorbed films.

\section{Computer simulations of the capillary wave spectrum of adsorbed films}

Computer simulations are an invaluable tool for the study of adsorption
phenomena \cite{binder11b}. Certainly, they have provided great
insight and a complementary perspective for the interpretation of different very
relevant experimental and theoretical findings. As a notable example, we can
mention studies on wetting and
prewetting transitions \cite{sikkenk87,finn88,finn89,sokolowski90} which parallel the first few
experimental reports \cite{rutledge92,cheng93}
performed  some years after the theoretical predictions \cite{ebner77}. Countless other
examples could be mentioned, but we will narrow this broad area and focus
merely on the study of capillary wave fluctuations of adsorbed films.

It is surprising to see that, despite the great number of studies on capillary
waves of free interfaces, only a few have been performed for the case of
adsorbed films. Similar to experimental realizations, most of such studies have
focused on  the analysis of capillary wave broadening. The findings reported
clearly indicate an interfacial roughness qualitatively in agreement with
$\Eq{ccwrsf}$, both as regards the expected decrease with increasing field strength
and the predicted increase with system size. However, a full quantitative
agreement has not been obtained, even though one has at hand both the upper
wave--vector cutoff, $q_{\rm max}$, and the intrinsic interfacial roughness
$\Delta_{\pi}$ as fitting parameters. Of particular interest is reference \cite{werner99}, where a
liquid--liquid interface of inmiscible polymer phases subject to van der Waals
forces was studied.  In this paper, it was indicated that the simulation results
could be only described if either,
i)  a film thick dependent interfacial width $\Delta_{\pi}$, or ii) a
film-thick dependent surface tension was incorporated into the classical theory.
Interestingly, these observations are  in  agreement with experimental
findings on a related system \cite{heilmann01}.

These studies notwithstanding, a detailed analysis of the full capillary wave 
spectrum of adsorbed films has not been performed until very recently
\cite{pang11,fernandez12,macdowell13}. This provides a very stringent test of
the classical theory, but requires dealing with two important difficulties
before the problem can be successfully tackled. Firstly, one needs to carefully
implement a practical methodology for defining the film profile that is
sufficiently robust to work also as the film thins. Secondly, it is required to
have adequate simulation techniques to asses independently the main properties
that are provided by the capillary wave spectrum, namely, the interface
potential and the surface tension. Let us now address each of these issues
briefly.

\subsection{Characterization of film profiles}

\label{sec:location}

Despite that the concept of ``surface'' is intuitive and familiar, the
mathematical problem of defining an interface from atomic scale data is 
surprisingly difficult and subject to arbitrariness. The problem is that our
intuition relies on the description of surfaces at large wavelengths, where the
continuity of such objects is not an issue.   In molecular simulations, however,
one deals with  sets of atomic positions, so that the data available is
essentially discrete. Thus, the only possible way out is to define a set of
criteria which will allow us to determine a smooth function $\ell(\rpar)$ from
the discrete atomic positions
\cite{benjamin92,werner97,chacon05,davidchack06,usabiaga09,zykova-tilman10,geysermans10}.

The selected criteria should provide  a mathematical surface, $\ell(\rpar)$,
that allows, on the one hand, to calculate a capillary wave spectrum for 
comparison with \Eq{ccws}, and on the other hand, to resolve the intrinsic density
profile $\rho_{\pi}(z)$ from the capillary wave fluctuations.

An apparently very simple procedure rooted on surface
thermodynamics is to divide the system into a set of $n$ elongated prisms of square basis and
fixed lateral area $A/n$ \cite{werner97,werner99,mueller00}. For prism $i$ one
calculates the lateral
average density $\rho_i(z)$ and defines the interfacial height
$\ell_i$ as the corresponding equimolar dividing  surface:
\begin{equation}
   \ell_i = \frac{1}{\Delta\rho_{\rm lv}} \int [ \rho_i(z) - \rho_v ] d z
\end{equation} 
where $\Delta\rho_{\rm lv}=\rho_l-\rho_v$.
This approach is simple to implement, has a thermodynamic basis and only one
arbitrary parameter, namely, the area of the prism's basis. In
principle, choosing a small lateral area one would achieve a high
resolution of the fluid interface, while simultaneously
suppressing the capillary wave fluctuations. Unfortunately, decreasing the
lateral area is at the cost of increasing the bulk fluctuations within
the prism's volume, which are of the order $\beta\rho\kappa V_i^{-1}$, with $V_i$ the prism volume. 
As a result,
the local dividing surfaces   $\ell_i$ pick up bulk like perpendicular fluctuations
that are unrelated to the interface position \cite{mueller00}. The side effects
of this coupling are that 1) a meaningful intrinsic density profile cannot be
extracted \cite{chacon05}, and 2) the spectrum of fluctuations at high
wave--vectors becomes strongly coupled to the bulk structure factor \cite{vink05}.
Despite these shortcomings, the surface tension can be still reliably extracted from
the spectrum, because it is obtained in the limit of small wave--vectors where the
finer details of the selected surface become irrelevant.

In order to obtain a more meaningful description of the interface, it is
required to abandon the dividing surface criterion and precisely pinpoint which
atoms actually lie on the interface. This task is very much facilitated when one
studies interfaces of strongly immiscible fluids \cite{benjamin92,luo06,geysermans10}. In
such cases, locating the highest molecules of the bottom phase and the lowest
molecules of the top  phase immediately allows us to define the surface with little
complications \cite{benjamin92,luo06,geysermans10}. For pure fluids, however,
surface atoms cannot be determined right away on the basis of their
perpendicular position, $z$. Rather, one needs to apply some additional criteria
to distinguish atoms of one phase from atoms of the other. For a
liquid--vapor system, this may be achieved by merely counting the number of 
neighbors of each atom, and attributing liquid--like character to those with
sufficiently close neighbors. For solid--liquid interfaces, with large
coordination number in both phases, more sophisticated criteria are required
\cite{davidchack06,zykova-tilman10,rozas11}.  Be as it may, once the atoms are
labeled as belonging to one phase or the other, the interface position can be
estimated as for the strongly segregated mixtures, using a simple height
criterion.

This procedure can be further refined using a Fourier description of the
surface as in \Eq{lfourier} \cite{chacon03,chacon05,chacon09}. 
Here, the height criterion is chosen for the purpose of selecting a few
roughly equally space ``pivot'' atoms. Then, Fourier components are determined in such a way as to
provide a surface with minimal area  going across the selected atoms.
Molecules close enough to this initial surface are incorporated into the set of
surface--pivots, and the procedure is iterated until a prescribed number of
pivots is achieved. In this way, the interfaces that are generated 
consistently have a fixed surface density. This method is certainly much more
time consuming than all others, and also demands large disk space.
However, it does indeed provide a capillary wave spectrum with the expected
monotonic increase, as well as highly structured intrinsic density profiles with
oscillatory behavior \cite{chacon03,chacon05,chacon09,bresme08,fernandez11b}.

\subsection{Calculation of interface potentials}

The interface potentials, or the related disjoining pressure, is the key
property for understanding adsorption phenomena. Experimentally, disjoining
pressures may be calculated using the captive bubble technique
\cite{blake72,blake75,vazquez05}, or Scheludko's method
\cite{scheludko67,langevin11}. Alternatively, one can estimate disjoining
pressures indirectly
by measuring the interfacial roughness as discussed previously
\cite{tidswell91b,doerr99,heilmann01, plech02}, or via the analysis of dewetting
patterns \cite{seemann01b,kim99}. However, essentially all of these methods are
limited to the study of relatively thick wetting films, and do not usually
probe the regime of very thin films.

Computer simulations offer the possibility to calculate interface potentials
reliably from essentially zero adsorption to the regime of thick wetting films
\cite{macdowell05,macdowell06,macdowell11,gregorio12}. With some additional reservations,
it even offers the possibility of extracting interface potentials in the range
 where adsorbed films are unstable. This issue is discussed at length
in a recent review, and will not be pursued further here \cite{macdowell11}.

The simulation setup that is usually employed consists of a tetragonal box of dimensions $L_x=L_y$  and
$L_z>L_x$. A substrate is placed at both sides of
the simulation box parallel to the x--y plane.
Performing a
grand canonical simulation at bulk coexistence, one fixes temperature, volume and chemical
potential, $\mu_c$.
In this way, a film consistent with the imposed thermodynamic conditions builds
on the substrate. However, because of the finite system size
of the simulation box, fluctuations away from the equilibrium state may be
observed (and enhanced in a controlled manner when necessary
\cite{berg92,fitzgerald99,mueller00,errington04,shen05}). This fact is
exploited in our procedure in order to measure free energy differences. During
the simulation, one simply monitors
the probability $P_{1/2}(N)$ of observing $N$ molecules inside the half of the
simulation box closest to the studied substrate
\cite{macdowell05,macdowell06}. 
Accordingly, one can define the instantaneous adsorption akin to that substrate
as:
 \begin{equation}
     \Gamma = ( N - \frac{1}{2} \rho_v L_x L_y L_z ) / L_x L_y
\end{equation}
A surface free energy or {\em effective interface potential} of a film with
adsorption $\Gamma$, or likewise, film thickness $\ell=\Gamma /
(\rho_l-\rho_v)$, can then be estimated up to additive constants as:
\begin{equation}
        g_{\mu_c}(\Gamma) = -\frac{k_B T}{L_x L_y} \ln P_{1/2}(\Gamma)
\end{equation}
Alternatively, one can map the interface potential into the average film
thickness $\ellav$ that is attained during simulations at constant $N$, and
hence, transform  $g_{\mu_c}(\Gamma)$ into  $g_{\mu}(\ellav)$. In
practice, at very low temperatures the vapor density is very low, the
bulk fluctuations are small and $\ellav$ as obtained from the criteria discussed
above is quite close to that obtained trivially from the mass balance
$\Gamma=\Delta\rho_{\rm lv}\,\ellav$.

This technique, which was first employed to study the wetting phase diagram of
polymers adsorbed on a brush \cite{macdowell05,macdowell06}, has been henceforth exploited
to great advantage by Errington and collaborators \cite{grzelak08,kumar11,rane11}. Indeed, once the
interface potential has been calculated, it provides a wealth of 
information on the wetting properties of the selected system, including the
order of the phase transition, the equilibrium adsorption, and the contact angle.
Particularly, it should be stressed that knowledge of the interface potential
at coexistence provides information on  interface potentials at whatever other
chemical potential, say $\mu'$, since one may be obtained from the other as a
Legendre transformation:
\begin{equation}
g_{\mu'}(\Gamma)=
g_{\mu_c}(\Gamma) + \Gamma (\mu' - \mu_c)
\end{equation} 
In this way, it is possible to map out the whole adsorption isotherm at the
chosen temperature \cite{macdowell06}.

\subsection{Recent simulation studies of the capillary wave spectrum}

\label{sec:simnew}

Very recently, we have performed a study of the capillary wave spectrum 
 that has allowed us to make a thorough test of the classical capillary wave theory 
\cite{macdowell13}.

The study was performed on 
 a well known model of Argon adsorbed on
solid Carbon Dioxide (ArCO$_2$) that has been employed in several studies of the prewetting
transition \cite{finn88,finn89,shi02,errington04}. In this model, Argon is described as
a truncated Lennard--Jones fluid, with energy parameter $\epsilon$,
 molecular diameter $\sigma$ and cutoff
distance $R_c=2.5\sigma$, while solid carbon dioxide is considered as
an inert flat wall including a long range tail
$V(z)=-\frac{\epsilon_w}{6\pi}z^{-3}$ with $\epsilon_w=65.56\epsilon\sigma^3$.
The simulation is performed at the established wetting transition of
$k_BT_w=0.60\epsilon$ \cite{errington04,gregorio12}.

This model has  many desirable features
that facilitate the analysis. Firstly, it exhibits a first order wetting
transition occurring at very low temperatures, close to the triple point of the
Lennard--Jones model \cite{errington04}. Because of the low temperature, bulk fluctuations are
small, and the interface can be identified with great accuracy using the
Intrinsic Sampling Method of Chacon and Tarazona \cite{chacon05}. The
Lennard--Jones fluid has truncated interactions, and is therefore short range.
As discussed at length in  section \ref{sec:dft},  this allows us to exploit a number of analytical 
results of Liquid State Theory in order to describe the fluid's behavior.
However, we still are mimicking at a reasonable degree of accuracy realistic
systems, since the fluid is subject to a long--range potential resulting from
the van der Waals interactions of the substrate's molecules.

\begin{figure}[t]
\centering
\includegraphics[clip,scale=0.30,trim=0mm 5mm 0mm 0mm,
angle=270]{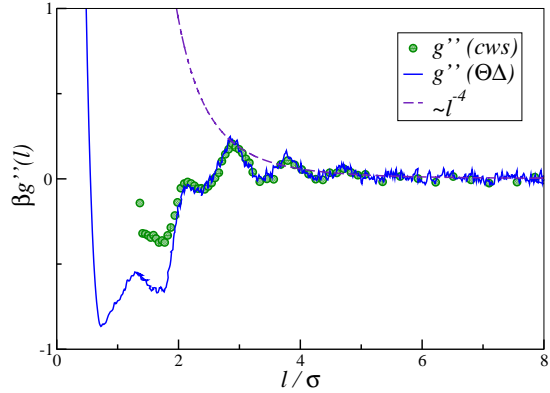}
\caption{Plot of the second derivative of the interface potential as obtained
from the capillary wave spectrum (symbols) and by numerical derivation of the
thermodynamic interface potential of Fig.~\ref{fig:ifp_and_pi} (full lines).  The
dashed lines are expectations from the Hamaker model of \Eq{hamakerz2} (results
adapted from Ref.\cite{macdowell13}, correspond to the same model and
temperature as in Fig.~4 and 5).
\label{fig:susceptibility}
}
\end{figure}

\begin{figure}[t]
\centering
\includegraphics[clip,scale=0.30,trim=0mm 5mm 0mm 0mm,
angle=270]{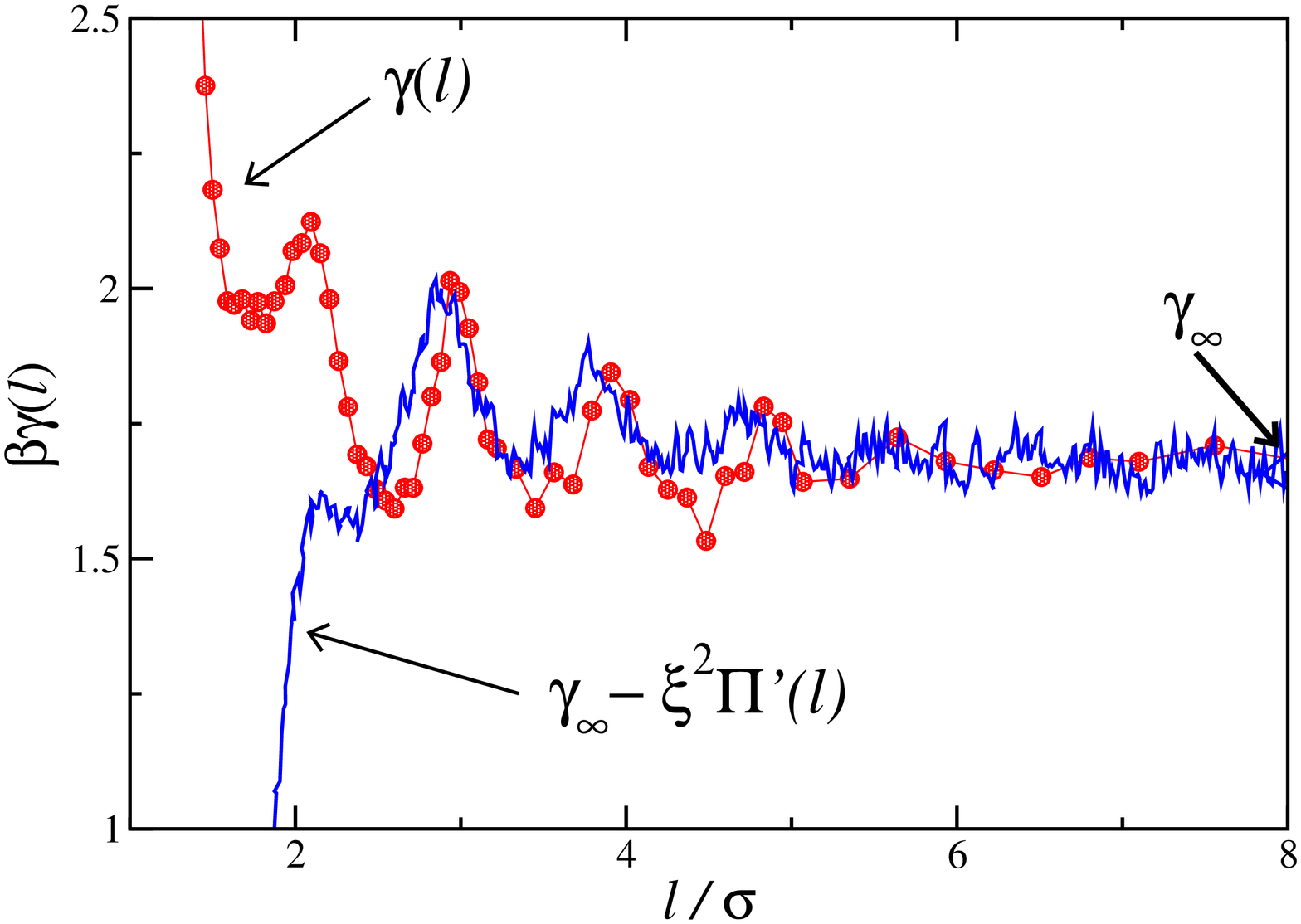}
\caption{Plot of the film--height--dependent surface tension  as obtained from
the capillary wave spectrum of adsorbed films (symbols) of different height $\ell$.
The lines are predictions from the model $\gamma_{\infty} - \xi_{\infty}^2
\Pi'(\ell)$, with $\Pi'(\ell)$ as obtained from numerical derivatives of the
disjoining pressure, Fig.~\ref{fig:ifp_and_pi} (results
adapted from Ref.\cite{macdowell13}, correspond to the same model and
temperature as in Fig.~4, 5 and 10.).
\label{fig:gamma}
}
\end{figure}

In our study, we simulated a large number of
systems with average film thickness ranging from one to ten molecular diameters.
For each configuration in a system, the Fourier components $\ell_{\qvec}$ 
of the film height profile were calculated and the thermal average
$<|\ell_{\qvec}|^2>$ was obtained. We then performed a fit of the form:
\begin{equation}\label{eq:cws}
   \frac{k_BT}{A<|\ell_{\bf q}|^2>} = g''_{\rm cws} + \gamma_{\rm
      cws}\, q^2 +
   \kappa_{\rm cws}\, q^4
\end{equation}
By comparing with expectations from the capillary wave theory, \Eq{ccwskappa}, the
coefficients $g''_{\rm cws}$, $\gamma_{\rm cws}$ and $\kappa_{cws}$, should provide
estimates of $g''$, $\gamma_{\infty}$ and $\kappa$, respectively.

Figure \ref{fig:susceptibility} shows the zero order coefficients  $g''_{\rm cws}$ as obtained from fits to the
capillary wave spectrum (symbols). The results are compared to the second
derivative of the interface potential described previously. The results indicate
an excellent agreement between both independent estimates. Interestingly, the
behavior is far from the asymptotic decay expected of the van der Waals forces,
and rather, exhibits now a very strong oscillatory behavior that is 
revealing the layered structure of the adsorbed films that was apparent 
in the density profiles of Fig.\ref{fig:profiles}.

The second order coefficient of the film, $\gamma_{\rm cws}$ is, according to
classical capillary wave theory, equal to the liquid--vapor surface tension,
$\gamma_{\infty}$, in all the range of film thickness. This expectation is tested
in Fig.\ref{fig:gamma}, where $\gamma_{\rm cws}$ is plotted as a function of $\ell$.
Clearly, we find that results for thick films provide an accurate
estimate of the surface tension as obtained independently for a free liquid
vapor interface, marked as a thick arrow on the figure. However, as the films get thinner, an oscillatory
behavior of  $\gamma_{\rm cws}$ becomes apparent.

This behavior cannot possibly be explained in the framework of classical
capillary wave theory, where the coefficient of the square gradient term is the
liquid--vapor surface tension essentially by definition. The question is
whether the interface potentials, or rather, the disjoining pressures that are
actually measured experimentally are able to cope alone with all the film--height dependency
required to describe the free energy of a rough interface as implied by
the definition of the Interface Hamiltonian Model.

In the next section we will review recent theoretical work addressing this
issue \cite{macdowell13}.

\section{An improved capillary wave theory}

\label{sec:ncwt}

Our simulation results of the capillary wave spectrum of adsorbed films
confirm the expectations of the classical theory for thick films. However, the
strong film--thick dependence of the surface tension that is observed for small
$\ell$ clearly indicates room for improvement.

Recently, we have suggested that a film--thick--dependent surface tension may be
explained by considering distortions of the intrinsic density profile beyond
the mere interfacial translations that are considered in the classical theory
\cite{macdowell13}.

The starting point is based on the seminal work of Fisher et al. on the
nature of the short--range wetting transition \cite{jin93,fisher94}. These authors
attempted to derive  the coarse--grained CWH from an
underlying microscopic density functional of the square--gradient type. Their
approach starts by
seeking  for a density profiles, $\rho(\rpar;\Sigma)$ that
extremalises the free energy functional subject to the constraint of a frozen
capillary wave, $\ell(\rpar)$, henceforth denoted as $\Sigma$ for short. 
The extremalised density profile, placed
back into the underlying functional yields a formal expression for the free
energy of the film of roughness $\Sigma$. This expression in then
compared with the CWH, and the appropriate interface potential and surface
tensions are identified.

Accordingly, let us assume that an adsorbed liquid film is frozen into a
configuration of fixed roughness $\Sigma$, and let $\rho(\rvec{}{};\Sigma)$ be the
corresponding average density. Within the double parabola
approximation, seeking for a solution of $\rho(\rvec{}{};\Sigma)$ amounts to solving
the Helmholtz equation, \Eq{helmholtz}, subject to the constraint  $\Sigma$. 

The capillary waves impose weak transverse perturbations on the
otherwise $z$ dependent density profile. Therefore, 
we suggest an expansion of $\rho(\rvec{}{};\Sigma)$ in transverse Fourier modes
as trial solution \cite{tolstov62}:
\begin{equation}\label{eq:trialq}
 \rho(\rvec{}{};\Sigma) = \rho_{\infty} + \sum_{\qvec{}{}}\Delta\rho_{\qvec}(z;\qvec{}{}) e^{i\qvec{}{}\cdot \rpar{}{}}
\end{equation} 
where, as in sections \ref{sec:lvi}--\ref{sec:afi}, $\rho_{\infty}$ denotes the asymptotic bulk
density, which is either, $\rho_l$ to the left of $\ell(\rpar)$ or $\rho_v$ to
its right; while we will assume for the time being a symmetric fluid ($b_l=b_v=b$) for the sake of
clarity. Later on, we will consider the more general solution that results when
$b_l\ne b_v$.

Notice that this trial solution
is of very general form.
Particularly, being expressed in terms of Fourier coefficients, it suggests from
the start that the density at a point $\rvec{}{}$ could depend, not just on the
local properties at that point, but rather on the structure of  the whole
interface. The need to account for such nonlocal effects, which is absent in the
theory of Fisher and Jin, has been strongly advocated by Parry and collaborators
\cite{parry06,parry07,bernardino09}.

It is also worth mentioning that an expansion of the density profile in Fourier
modes was previously employed by van Leeuwen and Sengers in their
study of interfaces under gravity as described by the Square--Gradient density
functional \cite{vanleeuwen89}. Relative to that work, however, we describe the
local free energy explicitly in the double parabola approximation. This
simplification will allow us to proceed without any further important
approximation and obtain results in closed form 
that provide a more transparent interpretation.

Coming back to the solution of \Eq{trialq}, we now apply the gradient operator twice on
$\Delta\rho(\rvec{}{};\Sigma)$, followed by substitution into \Eq{helmholtz}. It
is then found that the  transverse Fourier modes are
the solution of an ordinary second order differential equation:
\begin{equation}\label{eq:EDOq}
 \frac{d^2 \Delta \rho_{\qvec}(z)}{d z^2} - b_{q}^2\,
 \Delta\rho_{\qvec}(z) =  \frac{\beta}{C_{\infty}}V_{\qvec}(z)
\end{equation}
where $V_{\qvec}(z)$ are coefficients in a Fourier expansion of the external
field, while $b_{q}$ is now a wave--vector dependent correlation length,
promoting fast damping of small wavelength modes:
\begin{equation}
 b_{q}^2 = b_{\infty}^2 + q^2
\end{equation} 
Clearly, \Eq{EDOq} is formally equal to the equation for the independent
branches of  the planar liquid--vapor interface  that we discussed in section
\ref{sec:lvi}, and actually becomes identical to \Eq{EDO} for the special case $q=0$. 
Accordingly, the formal solution for the eigenfunctions, $\Delta\rho_{\qvec}(z)$
is  very similar and poses no difficulties. However, applying the
boundary conditions for the general case is here far more involved.

Since the solutions of \Eq{EDOq} for a structured adsorbed film
that we are seeking is rather lengthy, and could obscure the generalities of
the procedure, it will prove beneficial to consider first the capillary waves 
of a free liquid--vapor interface. 

\subsection{Liquid--vapor interface}

For a liquid--vapor interface we can ignore the external field  and obtain the
eigenfunctions exactly as was done previously for the planar interface. Solving
the homogeneous second order equation, followed by substitution of the Fourier
modes $\Delta\rho_{\qvec{}{}}(z)$ into the trial solution, \Eq{trialq}, we can
write:
\begin{equation}\label{eq:dpq}
\rho_{\rm lv}({\bf r}) = \rho_{\infty} + \sum_{\bf q} A_{\bf q}\, e^{\pm b_q z} e^{i{\bf
q}\cdot {\bf x}}
\end{equation} 
with the understanding that the liquid branch is obtained by setting
$\rho_{\infty}=\rho_l$ and $+b_q z$ as argument of the exponential; while the vapor branch
corresponds to $\rho_{\infty}=\rho_v$ and $-b_q z$ in the exponential function (c.f.
\Eq{piecewise}). Furthermore, owing to the piecewise structure of the double
parabola model, $\pm$ and $\mp$ signs will often appear in the expressions, and the
understanding is that the top sign refers to the liquid branch, while the bottom
sign corresponds to the vapor phase.

In order to obtain the Fourier coefficients of the
rough liquid--vapor interface, $A_{\bf q}$, we  evaluate $\rho({\bf r};\Sigma)$ at the boundary 
using the crossing criterion, such that the density at $z=\ell({\bf x})$ is
fixed to some prescribed value $\rho_{1/2}$. The result is:  
\begin{equation}
   \rho_{1/2} = \rho_{\infty} + \sum_{\bf q} A_{\bf q}\, f_{\qvec}(\ell({\bf
   x}))\,
  e^{i{\bf q}\cdot {\bf x}} 
\end{equation} 
where, for the sake of clarity, we have introduced the function
$f_{\qvec}=e^{\pm b_q z}$, with plus sign for the liquid branch and minus sign
for the vapor branch implied.

In order to proceed, we make a Taylor expansion of the right hand side up to
second order in powers of $\delta \ell(\rpar) = \ell(\rpar) - \ellav$:
\begin{equation}
 \Delta\rho_{1/2}= \sum_{\bf q} A_{\bf q} 
 \left \{ f_{\qvec}(\ellav) + f'_{\qvec}(\ellav)\,\delta \ell({\bf x}) +
 \frac{1}{2} f''_{\qvec}(\ellav)\, \delta \ell({\bf x})^2 \right \} 
    e^{i{\bf q}\cdot {\bf x}} 
\end{equation} 
where, as in section \ref{sec:lvi}, $\Delta\rho_{1/2}=\rho_{1/2}-\rho_{\infty}$. 

It is now convenient  to
collect the $\qvec\neq 0$ coefficients, of which we retain terms up to linear
order in  $\delta \ell(\rpar)$; separately from the  $\qvec=0$ term of which we
retain terms up to second order. We thus write:
\begin{equation}\label{eq:split}
\Delta\rho_{1/2}= S_0 + S_q
\end{equation}
with 
\begin{equation}
S_0 =  A_0 
 \left \{ f_0(\ellav) + f'_0(\ellav)\, \delta \ell({\bf x}) + 
     \frac{1}{2} f''_0(\ellav)  \delta \ell({\bf x})^2
\right \} 
\end{equation}
and: 
\begin{equation}
 S_q  =  \sum_{{\bf q}\ne 0} A_{\bf q}
 \left \{ f_{\qvec}(\ellav) + f'_{\qvec}(\ellav)\, \delta \ell({\bf x}) \right \} 
  e^{i{\bf q}\cdot {\bf x}}
\end{equation}
In order to obtain a relation for the coefficients $A_{{\bf q}\ne 0}$ we
make a Fourier transform on both sides of \Eq{split}, yielding: 
\begin{equation}
\Delta\rho_{1/2}\, \delta({\bf q}) =
  A_0 
 \left [ f_0(\ellav)\, \delta({\bf q}) + f'_0(\ellav)\, \delta \ell_{\bf q}
 \right ]
 + \sum_{{\bf q}'\ne 0} A_{\bf q'}  f_{\qvec'}(\ellav) \,
     \delta({\bf q}'-{\bf q})
\end{equation}
where we have neglected the last term in both $S_0$ and
$S_{q}$.  For ${\bf q}\ne 0$, all terms with a Dirac delta vanish, and we are
left with a relation between $A_{\bf q}$ and $A_0$:
\begin{equation}\label{eq:aqfa0}
A_{\bf q} f_{\qvec}(\ellav) = - A_0\, f'_0(\ellav)\, \delta \ell_{\bf q}
\end{equation}

For the ${\bf q}= 0$ term, we only need to make an unweighted lateral
average on both sides of \Eq{split}, which, retaining all terms in
$S_0$ and $S_q$, leads to the following equality:
\begin{equation}
\Delta\rho_{1/2} =  A_0 
 \left \{ f_0(\ellav)   + \frac{1}{2} f''_0(\ellav) \sum_{\qvec}    \delta \ell_{\bf q} \delta \ell_{-{\bf q}}
 \right \} + \sum_{\bf q} A_{\bf q}\, f'_{\qvec}(\ellav)\, \delta \ell_{-{\bf q}}
\end{equation}
Here, we have invoked once more Plancherel's theorem in order to write the
second term inside the brackets in terms of $\delta \ell_{\bf q} \delta
\ell_{-{\bf q}}$.

We can now eliminate the $A_{\qvec}$ coefficients of the above result using
\Eq{aqfa0} and obtain a closed expression for $A_0$ in terms of $\delta \ell_{\bf q}$:
\begin{equation}
\Delta\rho_{1/2} =   A_0\,  f_0(\ellav) \left \{ 
 1  + \sum_{\qvec} \left [  \frac{1}{2} \frac{f''_0(\ellav)}{f_0(\ellav)} - 
                     \frac{f'_0(\ellav)}{f_0(\ellav) }
			   \frac{f'_{\qvec}(\ellav)}{f_{\qvec}(\ellav)}
			  \right ]
\delta \ell_{\bf q}\delta \ell_{-{\bf q}}
\right \}
\end{equation}

Using the results of the last two paragraphs, we readily obtain the
following expression for the coefficients of the series:
\begin{equation}
 A_0 =  \Delta\rho_{1/2}\, f_0^{-1}(\ellav) \left \{
    1 -  \sum_{\bf q} \left [ \frac{1}{2} \frac{f''_0(\ellav)}{f_0(\ellav)} -
       \frac{f'_0(\ellav)}{f_0(\ellav) } \frac{f'_{\qvec}(\ellav)}{f_{\qvec}(\ellav)}
      \right ]
    \delta \ell_{\bf q}\delta \ell_{-{\bf q}}   
  \right \}
\end{equation}
\begin{equation}
 A_{\bf q} = - \Delta\rho_{1/2}\, 
 \frac{f'_0(\ellav)}{f_0(\ellav)\, f_{\qvec}(\ellav)}\, \delta \ell_{\bf q}
\end{equation}
We can now replace the coefficients back into  \Eq{dpq}, and obtain the sought
equation for the liquid--vapor density profile of a rough interface:
\begin{equation}\label{eq:lviq}
\begin{array}{lll}
 \rho({\bf r};\Sigma) & = & 
\rho_{\pi}(z;\ellav) 
  + \frac{1}{2} \Delta\rho_{1/2} e^{\pm b(z- \ellav)} \sum_{\bf q}  \left [ b^2 + q^2 \right ]
  \delta \ell_{\bf q}\delta \ell_{-{\bf q}}   \\
   & & \\
  & &  \mp b\, \Delta\rho_{1/2} \sum_{\bf q}  e^{\pm b_q (z- \ellav)}\, \delta \ell_{\bf q}\, e^{i{\bf q}\cdot {\bf x}}
\end{array}
\end{equation} 
where $\rho_{\pi}(z;\ellav)$ here stands for the intrinsic density profile of a
liquid--vapor interface in the double parabola approximation, \Eq{dplvcc}.
This result already without any further elaboration provides us with a great deal of
insight on how capillary--waves modify the structure of an interface. For the
time being,  we briefly mention here just a few:
\begin{itemize}
\item The perturbed density profile $\rho({\rvec{}{}};\Sigma)$, has a leading
order contribution equal to the intrinsic density profile of the planar
interface.
\item To linear order in the perturbation, the capillary waves provoke an
additional dependence on the transverse direction, $\rpar$. Contrary to
expectations from the classical theory, however, the decay of this perturbation
is not merely given by the inverse correlation length, $b$. Rather, short
wavelength perturbations decay at a faster, wave--vector dependent rate, $b_q$.
\item To second order in the perturbation, the capillary waves result in an
additional $z$ dependent contribution to the density profile which is
responsible for the capillary wave broadening. Following expectations from the
classical theory, the broadening depends on the average squared amplitude of the
capillary perturbation, $\sum |\delta \ell_{q}|^2$ (c.f. \Eq{rhobexp}). However, as first noted in
\cite{macdowell13}, there is an additional capillary wave broadening mechanism
of order $\sum | q\,\delta \ell_{q}|^2$ not included in the classical theory.  
\end{itemize} 
We postpone further discussion of the implications to a later section, and
consider now the density profile of an adsorbed film.

\subsection{Adsorbed films}

Having studied the case of a free liquid--vapor interface, let us now consider
the role of capillary waves on the structure of an adsorbed film. Clearly, by
definition the capillary waves propagate at the liquid--vapor interface, so the
question is, to what extent do they penetrate and distort the structure of the layered
film. The results of Fig.\ref{fig:rhorough}, are quite revealing in this regard, and show that,
despite the obvious broadening of the density profile at the liquid--vapor
interface, the layering structure of the adsorbed liquid remains essentially unchanged as the system
size increases, exhibiting no sign of broadening what so ever. This suggests that we can employ 
the superposition approximation of section \ref{sec:afi}, with all of the capillary 
wave effects lumped into the liquid--vapor density profile. Accordingly, we consider solutions of the form
(c.f. \Eq{superpos0}):
\begin{equation}
  \rho(\rvec{}{};\Sigma) = [ 1 + h_{\rm wl}(z) ]\, \rho_{\rm lv}(\rvec{}{};\Sigma)
\end{equation} 
where, at this stage, $\rho_{\rm
lv}(\rvec{}{};\Sigma)$ adopts the general form for the solution of the Helmholtz
equation, as given in \Eq{dpq}.

In order to look for the coefficients of the above equation, we employ again the
crossing criterion, and set:
\begin{equation}
 \rho_{1/2} = \left [ 1 + h_{\rm wl}(\ell(\rpar)) \right ] \times
              \left [ \rho_{\infty} + \sum_{\bf q} A_{\bf q} e^{\pm
   b_q \ell({\bf x})} e^{i{\bf q}\cdot {\bf x}}
               \right ]
\end{equation} 
Calculations are now much more lengthy than before, but proceed exactly in the same
manner. i.e, a Taylor expansion in powers of $\delta\ell$ is performed, the
resulting expressions are Fourier transformed, and the set of linear equations
is solved. We omit the lengthy details here and write the final solution, 
which can be given in compact form in terms of an intrinsic density profile and
recovers the result for the liquid--vapor interface  (c.f. \Eq{lviq}) when
$h_{\rm wl}(z)=0$:
\begin{equation}\label{eq:rhoroughnew}
\begin{array}{lll}
\rho({\bf r};\Sigma) &  = &
\rho_{\pi}(z;\ellav) + \frac{1}{2} \sum_{\bf q} \left [
   \frac{d ^2 \rho_{\pi}(z;\ellav)}{d  \ell^2} \mp
      \frac{d  \rho_{\pi}(z;\ellav)}{d  \ell} \frac{q^2}{b}
  \right ] | \delta \ell_{\bf q}|^2
   \\
    & & \\
    & &
      +  \frac{d  \rho_{\pi}(\ellav;\ellav)}{d  \ell}
	\sum_{\bf q} \delta \ell_{\bf q} \; (1 + h_{\rm wl}(z))
  e^{\pm b_q (z- \ellav)} e^{i{\bf q}\cdot {\bf x}}
  \end{array}
\end{equation} 
where now, the intrinsic density profile is that given for planar adsorbed films
in section \ref{sec:afi}, \Eq{superpos}.

This equation shares the same qualitative implications that where found for the
free liquid--vapor interface. Of particular interest is the laterally averaged
density profile, which is immediately recovered from the above result as the
$q=0$ Fourier coefficient:
\begin{equation}\label{eq:ncwb}
\rho(z;\Sigma)   =
\rho_{\pi}(z;\ellav) + \frac{1}{2} \sum_{\bf q} \left [
   \frac{d ^2 \rho_{\pi}(z;\ellav)}{d  \ell^2} \mp
       \frac{d  \rho_{\pi}(z;\ellav)}{d  \ell} \frac{q^2}{b}
	  \right ] | \delta \ell_{\bf q}|^2
\end{equation}
if we now recall that $\sum_{\bf q}  | \delta \ell_{\bf q}|^2 = \langle\delta
\ell^2\rangle_{\rpar}$, while $\sum_{\bf q}  | q \delta \ell_{\bf q}|^2 = \langle\nabla 
\ell^2\rangle_{\rpar}$, we see that the rough interface profile may be written as an
expansion in powers of the squared amplitudes {\em and gradient}:
\begin{equation}\label{eq:ncwb2}
 \rho(z;\Sigma)   =
 \rho_{\pi}(z;\ellav) + \frac{1}{2} 
    \frac{d ^2 \rho_{\pi}(z;\ellav)}{d  \ell^2} \langle\delta \ell^2\rangle_{\rpar} 
  \mp \frac{1}{2} \frac{d  \rho_{\pi}(z;\ellav)}{d  \ell} \langle(\nabla \ell)^2\rangle_{\rpar}
\end{equation} 
where $\langle \rangle_{\rpar}$ denotes the unweighted lateral average $1/A\int d\rpar$.

In order to work out the implications of this equation more transparently,
it is convenient
to consider the limit of thick adsorbed films, where the wall correlations have
essentially died out, so that $h_{wl}(z)=0$.  In that case, the liquid--vapor
interface is unperturbed by the substrate and is equal to the free liquid--vapor
interface. Using now \Eq{piecewise}, it can be readily seen  that
$d^2 \rho_{\pi}(z)/d\ell^2$ and $\mp b d  \rho_{\pi}(z)/d  \ell$ become equal within 
the double parabola approximation, and the capillary wave perturbed density
profile becomes:
\begin{equation}\label{eq:ncwb3}
 \rho(z;\Sigma)   =
 \rho_{\pi}(z;\ell) + \frac{1}{2} 
    \frac{d ^2 \rho_{\pi}(z;\ellav)}{d  \ell^2} \left \{ \langle\delta \ell^2\rangle_{\rpar} 
     +
  \xi^2_{\infty}  \langle (\nabla \ell)^2\rangle_{\rpar} \right \}
\end{equation} 
Notice that this equation is similar to the classical result for the density
broadening due to capillary waves (c.f. \Eq{rhobexp}), however, an extra term of order
square gradient as identified in Ref.\cite{macdowell13} for the first time is present.

\subsection{Capillary wave spectrum}

Having obtained the density profile consistent with the constraint $\Sigma$,
we can now plug back $\rho({\bf r};\Sigma)$ into the
free energy functional in order to estimate the free energy of the assumed capillary
wave fluctuation.
In practice,
because we are interested in films subject to long--range van der Waals forces,
it will suffice, as was the case in section \ref{sec:lvs}, to evaluate the dominant
external field contribution. Luckily, since we are assuming an external field
that only depends on the perpendicular direction, only the laterally averaged
density profile is required. Hence, using \Eq{ncwb3} into \Eq{gext}, readily 
yields:
\begin{equation}
g_V(\Sigma) = g_V(\ellav) +  \frac{1}{2} 
        \left \{ \langle\delta \ell^2\rangle_{\rpar}
	      +
		  \xi^2_{\infty} \langle (\nabla \ell)^2\rangle_{\rpar} \right \}  
\int   \frac{d ^2 \rho_{\pi}(z)}{d  \ell^2} V(z) d z
\end{equation} 
In order to obtain the overall free energy of the perturbation, we now replace
this equation into  \Eq{cwhlin}, using $g_V(\Sigma)$ as an estimate for the full
$g(\Sigma)$. Comparing the result with \Eq{pr1},
we see that the term of order $\langle\delta \ell^2\rangle_{\rpar}$ can be readily
identified with $g''_V(\ellav)$. However, the  term
of order $ \langle\nabla \ell^2\rangle_{\rpar}$ that is proportional to the surface area of
the capillary wave, can only be incorporated as an effective contribution to
the surface tension. Whence, we can write, in a more compact form:
\begin{equation}\label{eq:gvq}
 g_V(\Sigma) = g_V(\ellav) + \frac{1}{2} g_V''(\ellav) \langle\delta \ell^2\rangle_{\rpar} 
       + \Delta\gamma(\ellav) \langle\nabla \ell^2\rangle_{\rpar}
\end{equation} 
where the coefficient $g_V''(\ellav)$ is given
exclusively in terms of the intrinsic density profile:
\begin{equation}
  g_V''(\ell)  =  \int \frac{d ^2 \rho_{\pi}(z)}{d  \ell^2} V(z) dz
\end{equation} 
while the $\ell$ dependent contribution to the surface tension is
\cite{macdowell13}:
\begin{equation}\label{eq:deltagsym}
    \Delta\gamma(\ell) = \xi^2_{\infty}\,  g_V''(\ell)
\end{equation} 
Substitution of \Eq{gvq} into \Eq{cwhlin}, and transforming fluctuations of
$\ell(\rpar)$ as
usual, we finally obtain, for the Fourier transformed Hamiltonian, the following
result:
\begin{equation}\label{eq:ncwhfour}
 H[\Sigma] = A g(\ellav) + \frac{1}{2} \sum_{\qvec} [ g''(\ellav) +
    ( \gamma_{\infty} + \Delta\gamma(\ellav)   )\, q^2 ]
      \delta\ell_{\qvec} \delta\ell_{-\qvec}
\end{equation} 
Comparing this result with \Eq{cwhfour}, we readily see that the consistent
statistical thermodynamic treatment of the capillary wave Hamiltonian 
provides a 
free energy very much in agreement with the phenomenological classical capillary wave theory.
The difference is, however, that the square--gradient coefficient, corresponding to
$\gamma_{\infty}$ in the classical theory, is augmented with an extra $\ell$
dependent contribution $\Delta\gamma(\ell)$.

We can now readily obtain the  spectrum of fluctuations consistent with
\Eq{ncwhfour}, by following the same procedure discussed earlier (section \ref{sec:ccws}). The result
is:
\begin{equation}\label{eq:ncws}
  \frac{k_BT}{A\langle|\ell_{\qvec}|^2\rangle} = g''(\ell) +  ( \gamma_{\infty} +
  \Delta\gamma(\ell) ) q^2
\end{equation} 
This equation shows that the $q^2$ coefficient of the CWS is not just
$\gamma_{\infty}$ as implied by the classical theory,  \Eq{ccws}, but rather,
picks up an additional $\ell$ dependent contribution, which, under some
simplifying assumptions follows \Eq{deltagsym}.

Since both $g''(\ell)$ and $\gamma(\ell)$
are available from that study of Ref.\cite{macdowell13}, and  $g''(\ell)$ is essentially dominated by
$g_V''(\ell)$, is suffices to employ $\xi^2_{\infty} $ as an empirical parameter in order
to test the expectation of the above equation, namely, that the coefficient of
order $q^2$ in the CWS, obeys $\gamma(\ell)=\gamma_{\infty} + \xi^2_{\infty}  g''(\ell)$. 
The result of the comparison, depicted in Fig.\ref{fig:gamma}, shows clearly a 
strong correlation between $\gamma(\ell)$ and $g''(\ell)$, in all the range
where $g''(\ell)$ is dominated by the long--range forces. The agreement clearly
breaks down below two molecular diameters but in that regime the short range forces
become relevant and \Eq{deltagsym} need not hold any longer. 

Particularizing \Eq{ncwhfour} to the specific case of an algebraically decaying
potential, we recover a   result obtained previously from the nonlocal theory
of interfaces \cite{bernardino09}. That theory has been largely applied  to the
study of short range wetting, but its implications as regards to capillary wave
broadening, and the effect of long range forces have hardly been considered
\cite{bernardino08,bernardino09}. An effort is still needed to assess to what
extent are both approaches equivalent. 

Before closing, let us mention that, if we explicitly consider the asymmetry of
the liquid and vapor phases,  $\Delta\gamma(\ell)$ then picks up additional
terms of order $g'(\ell)$ which vanish in the limit of a symmetrical fluid, as
noted previously \cite{bernardino09}. A result exact for the double parabola
model up to linear order in the asymmetry is:
\begin{equation}
   \Delta\gamma(\ell) = \frac{1}{2} ( \xi_l^2 + \xi_v^2 ) g''(\ell) 
    -  \frac{1}{4} ( \xi_l^2 - \xi_v^2)(\frac{1}{\xi_l} + \frac{1}{\xi_v})
	 g'(\ell)  + O(( \xi_l^2 - \xi_v^2)^2 )
\end{equation} 
According to the analysis
performed in \cite{macdowell13}, terms governed by $g'(\ell)$ do not yield any
significant improvement for the model of Argon on carbon dioxide, indicating a small role of 
the asymmetry even at such low temperature.

\subsection{Capillary wave broadening}

Another important implication of the theoretical analysis of 
Ref.\cite{macdowell13} refers to the roughening of the average interface profile
as described by \Eq{ncwb2} and \Eq{ncwb3}. Clearly, the
first two terms on the right hand side of \Eq{ncwb2}, are exactly as predicted by the
classical capillary wave theory for an expansion of $\rho(\rvec{}{};\Sigma)$
about $\ellav$ (c.f. \Eq{rhobexp}), but the third term implies a contribution of
order square--gradient to the capillary wave broadening, that had not been
identified previously.  Physically, this contribution implies that the capillary waves do not merely propagate by
translation of the original perturbation, $\ell(\rpar)$, but rather, are
distorted due to curvature of  the interface, and lead to iso--density lines
that are no longer parallel. 

In order to assess the extent of the square--gradient contribution it is more
convenient to consider the simplified  \Eq{ncwb3}. Comparing that equation with
\Eq{rhobexp}, shows that the  the combined contributions of translation and distortion
to capillary wave
broadening may be interpreted as an effective  translation enhanced by terms of
order square--gradient. Whence, the combined effects may be lumped into a single
roughening parameter $\Delta^2_{\rm cw}$, as:
\begin{equation}
\begin{array}{ccc}
  \Delta^2_{\rm cw} & = &  \langle\delta \ell^2\rangle + \xi^2_{\infty} \langle(\nabla \ell)^2\rangle  \\
   & & \\
   & = & \sum_{\qvec} ( 1 + \xi^2_{\infty} q^2 ) \langle|\delta \ell_{\qvec}|^2\rangle
\end{array}
\end{equation} 
where the Fourier amplitudes, $\delta\ell_{\qvec}$ are given by \Eq{ncws}. 
This sum may be transformed
into an integral as for the classical theory. For the sake of generality,
however, we extend the result of \Eq{ncws}, by adding the next to leading order
correction, $\kappa\,q^4$ to  the denominator of $\langle|\delta \ell_{\qvec}|^2\rangle$, as implied by the Helfrich
Hamiltonian and discussed at length by Mecke \cite{mecke01} (c.f.
\Eq{ccwskappa}). Performing the integral to leading
order in $\kappa$, we obtain an expression in real algebra that smoothly transforms into the
classical result, \Eq{ccwr}:
\begin{equation}\label{eq:ncwrgt}
 \Delta_{\rm cw}^2 =  \Delta_{\rm \gamma}^2 +  \Delta_{\rm \kappa}^2
\end{equation} 
with
\begin{equation}\label{eq:ncwrg1}
  \Delta_{\gamma}^2 = \frac{k_B T}{4\pi\gamma(\ell)} \left [
  \frac{ \xi^2_{\parallel} - \xi^2_{\infty} }{\xi^2_{\parallel} - 2\xi^2_{\rm
R}} \right ] \ln \left (
  \frac{ 1 + \xi_{\parallel}^2 q_{\rm max}^2}{1 + \xi_{\parallel}^2 q_{\rm
min}^2 } \right )
\end{equation} 
and 
\begin{equation}\label{eq:ncwrg2}
 \Delta_{\kappa}^2 =  \frac{k_B T}{4\pi}\frac{\xi^2_{\infty}}{\kappa} 
\left [   \frac{\xi_{\parallel}^2  - ( 1 + \xi_{\parallel}^2/\xi^2_{\infty} ) \xi^2_{\rm R}
   }{ \xi_{\parallel}^2  - 2 \xi^2_{\rm R}  } \right ] \ln \left (
   \frac{ \xi_{\parallel}^2  - ( 1 - \xi_{\parallel}^2 q_{\rm max}^2 ) \xi^2_{\rm R}  }
        { \xi_{\parallel}^2  - ( 1 - \xi_{\parallel}^2 q_{\rm min}^2 )
\xi^2_{\rm R}  } \right )
\end{equation} 
In these equations, $\gamma(\ell)$ refers to the $\ell$ dependent quantity
$\gamma(\ell)=\gamma_{\infty} + \xi_{\infty}^2 g''(\ell)$ defined previously;  
$\xi_{\infty}$ is the bulk correlation length; a sort of
bending correlation length $\xi_{\rm R}^2 = \kappa/\gamma$ emerges naturally;
while the parallel correlation length is now given as:
\begin{equation}
\begin{array}{ccc}\label{eq:newparallel}
\xi^2_{\parallel} & = & \frac{\gamma(\ell)}{g''(\ell)} \\
   & & \\
                  & = & \frac{\gamma_{\infty}}{g''(\ell)} + \xi_{\infty}^2
\end{array}
\end{equation} 

Note that \Eq{ncwrgt} was written in terms of the two contributions $\Delta_{\gamma}$ and $\Delta_{\kappa}$ for the sake of
brevity, but physically, it is more relevant to write
\begin{equation}
   \Delta^2_{\rm cw} = \Delta^2_{\rm tras} + \Delta^2_{\rm dis}
\end{equation} 
where $\Delta^2_{\rm tras}=\langle\delta\ell^2\rangle$ is the interface translation 
roughening, while $\Delta^2_{\rm dis}=\xi^2_{\infty} \langle(\nabla \ell)^2\rangle$ is the
interface distortion roughening. These contributions may be readily
recognized from \Eq{ncwrg1}--\Eq{ncwrg2} by noticing that $\Delta^2_{\rm dis}$ has a linear prefactor of order
$\xi_{\infty}^2$.

The picture that emerges from these equations is that the roughness of an
interface as measured from experiments is, for the most general case, a very
subtle  phenomenon involving bulk and interface properties as well as surface
interactions. It is relevant to recall in this context, that the above result, actually, 
was obtained via the simplifying assumption of i) a symmetric fluid, and ii) the
liquid--vapor interface unaffected by the field.

In practice, the expression obtained  here does not upset the good
experimental agreement of the classical theory in the low field limit. Indeed,
assuming $\xi_{\parallel}$ is larger than the system size, $\xi_{\parallel}q_{\rm
min}\gg 1$, and neglecting $\xi_R$, we find:
\begin{equation}
  \Delta^2_{cw} = \frac{k_B T}{2\pi\gamma_{\infty}} \left [
     \ln \left ( \frac{q_{\rm max}}{q_{\rm min}} \right ) + \frac{1}{2} \xi_{\infty}^2 q_{\rm max}^2 \right ]
\end{equation} 
whence, the distortion contribution to the interface is a constant and does not
upset the $\ln L$ dependence observed experimentally. Presumably, the extra term
could be tested experimentally by performing scattering experiments with
different instrumental cutoffs.

If, however, we consider the strong field limit, we find that indeed, because of
the $\ell$ dependence of the surface tension, the predicted roughness exhibits
important differences with the classical theory, already under the simplifying
assumption of vanishing $\xi_R^2$:
\begin{equation}\label{eq:sfncwt}
  \Delta^2_{cw} =  \frac{k_B T}{2 \pi \gamma(\ell)} \left [
  \left ( 1 - \frac{\xi_{\infty}^2}{\xi_{\parallel}^2 } \right ) \ln \left (
\xi_{\parallel} q_{\rm max} \right )
  +  \frac{1}{2} \xi_{\infty}^2 q_{\rm max}^2 \right ]
\end{equation} 
Particularly, this equation predicts a sharper increase of the roughness towards the low field limit
than the classical theory, and such behavior does indeed conform qualitatively
to some of the experimental findings \cite{wang99,heilmann01,plech02}.

The results above await verification, but it is interesting to point out that
already some authors  have noticed that their
results could be better described with a film--height dependent surface tension
\cite{werner99,wang99,heilmann01,plech02}.

Particularly relevant here is the work by Wang et al. \cite{wang99}, who studied
thin polymer films at conditions where the bulk fluid exhibits liquid like
behavior. Unexpectedly, the specular and diffuse x--ray scattering experiments
performed could not be described simultaneously from the classical capillary
wave theory of section \ref{sec:roughness}. On the one hand, their diffuse scattering experiments 
could only be described with an enhanced surface tension. On the other hand,
such enhancement would result in interfacial roughness much smaller than implied
by their specular reflectivity data.  The authors reconciled these
conflicting observations by hypothesizing a strong viscoelastic behavior of the
polymer films, implying some degree of ``glassification'' as a result of
confinement. However,  \Eq{sfncwt} shows that, due to the additional contribution
$\xi_{\infty}^2 q_{\rm max}^2$,  an enhanced surface tension is actually compatible with 
an enhanced roughness, and could also serve as an explanation for the observed behavior.

It is tempting to apply the result above to the case of a fluid--fluid
interface  subject to gravity. As the fluid approaches the critical point,
$\xi_{\parallel}$ approaches $\xi_{\infty}$ (c.f. \Eq{newparallel}). Whence, 
according to \Eq{sfncwt}, the logarithmic contributions of translation and distortion 
roughening cancel each other exactly, and only an ultraviolet cutoff term 
approaching $q_{\rm max}^2$ survives.
This result is in agreement with  theoretical estimates by van Leeuwen and Sengers, who
considered the role of capillary wave translations on the Fisk--Widom intrinsic
density profile and predicted a negligible roughening at criticality \cite{vanleeuwen89}. 
Interpertations based on \Eq{sfncwt} must be taken with some caution, however, since our
approach assumes the external field does not perturb the intrinsic density
profile. As
is clear from \Eq{generalbar}, this is a reasonable approximation for an
incompressible fluid, but breaks down at criticality because of the diverging 
compressibility.

\subsection{Discussion}

\begin{figure}[t]
\centering
\includegraphics[clip,scale=0.30,trim=0mm 5mm 0mm 0mm,
angle=270]{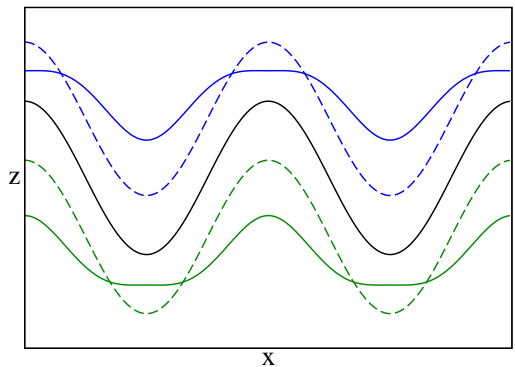}
\caption{Sketch with iso--density lines representing a monochromatic capillary
wave perturbation. The black line is the initial perturbation as imposed by the
crossing criterion, satisfying $\rho(\rpar;z=\ell(\rpar))=\rho_{1/2}$. The blue
lines (top) correspond to iso-density curves inside the vapor phase with
$\rho=\frac{1}{2}(\rho_v -\rho_{1/2})$, while the green curves (bottom)
correspond to iso-density curves in the liquid phase, $\rho=\frac{1}{2}(\rho_l -\rho_{1/2})$.
Full lines are predictions from \Eq{dpm}, while dashed lines are the classical
theory in the double parabola approximation (i.e., \Eq{dpm} with gradient and
Laplacian terms ignored). The monochromatic perturbation has wavelength of five
and amplitude 9/10 in units of the correlation length.
\label{fig:sketch_iso}
}
\end{figure}

In this section we will discuss some physical implications of the general
result, \Eq{rhoroughnew}, leading naturally to the identification of the
film--height dependent surface tension. 

In that equation, the $q=0$ contribution of $\rho(\rvec{}{};\Sigma)$ features terms
of order $|\delta \ell_{\qvec}|^2$ and $|q\,\delta \ell_{\qvec}|^2$, while, due to mathematical difficulties, 
$q\ne 0$ Fourier coefficients are given only to order $\delta \ell_{\qvec}$. 
The $q=0$ mode of  \Eq{rhoroughnew} is written in convenient
form in \Eq{ncwb2}. There, it is clear that, apart from the intrinsic density
profile, there appear terms that are lateral averages of $\delta\ell(\rpar)$
and $(\nabla\ell)^2$. 
Since the $q=0$ mode in a Fourier expansion is
essentially an unweighted lateral--average of the full solution,
we conclude that $\rho(\rvec{}{};\Sigma)$
must feature terms that are of order $\delta\ell(\rpar)^2$,
and $(\nabla\ell(\rpar))^2$ that are missing in \Eq{rhoroughnew} because of the
truncation to order $\delta\ell_{\qvec}$ in the $q\ne 0$ Fourier modes. 

A further important mathematical feature of the solution may be noticed by
expanding the exponential function of \Eq{rhoroughnew}, which immediately shows
that $\rho(\rvec{}{};\Sigma)$ also features terms of order
$q^2\,\delta\ell_{\qvec}$
which can be immediately related to the Laplacian, $\nabla^2\ell(\rpar)$.

Thus, to the order that is implied in \Eq{rhoroughnew}, we can conclude that
$\rho(\rvec{}{};\Sigma)$ may be expressed as a function of  $\delta\ell(\rpar)$,
$[\nabla\ell(\rpar)]^2$ and $\nabla^2\ell(\rpar)$. Whence, the strong
nonlocality that is potentially  implied in the trial solution of \Eq{trialq},
is to this order a ``weak'' nonlocality, such that $\rho(\rvec{}{};\Sigma)$ may be
expressed as a local function of the interface displacement, squared gradient
and Laplacian. This feature of the solutions of the Helmholtz equation were
already noticed some time ago \cite{fisher94}, but not explored any further.
The weak nonlocality implied here, however, provides a much more complex
behavior than the classical theory, where $\rho(\rvec{}{};\Sigma)$  is only a local
function of $\delta\ell(\rpar)$.

Heuristically, we can suggest the following extended local form for the rough
liquid--vapor density profile $\rho(\rvec{}{};\Sigma)$ in the double parabola
approximation \cite{macdowell12}:
\begin{equation}\label{eq:dpm}
\rho(\rvec{}{};\Sigma)) = \left \{
\begin{array}{ll}
 \rho_l - \Delta \rho_{1/2} e^{(L_{\bf x}+b_{\bf x})(z-\ell)} & z<\ell(\rpar) \\
 &  \\
 \rho_v + \Delta \rho_{1/2} e^{(L_{\bf x}-b_{\bf x})(z-\ell)} & z > \ell(\rpar)
\end{array}
\right .
\end{equation}
where $L_{\bf x}=\frac{1}{2}(1+(\nabla\ell)^2)^{-1}\nabla^2\ell$ and
\begin{equation}\label{eq:bx}
 b_{\bf x} =  \frac{\sqrt{b^2+ b^2(\nabla \ell)^2 + \frac{1}{4}(\nabla^2
\ell})^2}
                        {1+(\nabla \ell)^2}
\end{equation}
is a local, curvature dependent correlation length that plays the same role as
$b_q$ in the Fourier mode theory. The motivation for this result may be grasped
intuitively for the special case of a film profile $\ell(\rpar)$ locally
exhibiting finite
gradient but zero curvature $\nabla^2\ell(\rpar)=0$. 	In that case, \Eq{dpm} is
transformed into a function $\rho_{\pi}(h(z,\rpar))$  of the single variable:
\begin{equation}
 h =     \frac{z-\ell(\rpar)}{\sqrt{1 + (\nabla\ell)^2}}
\end{equation} 
whence, in contrast with the classical theory, which assumes vertical interface
translations $z\to z-\ell(\rpar)$ of the profile, the above result considers the
density of the inclined profile as given by the shortest  (perpendicular) distance to the
interface (rather than the vertical distance). This {\em ansatz}, which seems rather natural on physical grounds, has been
invoked occasionally in improved capillary wave models for the description of the liquid--vapor
interface \cite{mecke99b,blokhuis09}.

In order to compare the result of \Eq{dpm} with that of \Eq{rhoroughnew}, we expand
$\rho(\rvec{}{};\Sigma))=f(\rvec{}{};\delta\ell,(\nabla\ell)^2,\nabla^2\ell)$
about the planar interface, yielding:
\begin{equation}
\begin{array}{ccc}
 \rho(\rvec{}{};\Sigma)) &  = & \rho_{\pi}(z) +
\frac{d \rho_{\pi}}{d\ell}\delta\ell(\rpar) \mp \frac{1}{2}
\frac{z-\ell_{\pi}}{b} 
 \frac{d \rho_{\pi}}{d\ell} \nabla^2\ell(\rpar)  
+ \frac{1}{2} (z-\ell_{\pi})
  \frac{d \rho_{\pi}}{d\ell} (\nabla\ell(\rpar))^2  \\
 & & \\
 & &
 + \frac{1}{2} \frac{d^2 \rho_{\pi}}{d\ell^2} \delta\ell(\rpar)^2 +
O(\delta\ell\nabla^2\ell,\delta\ell^3,(\nabla\ell)^2,(\nabla^2\ell)^2)
\end{array}
\end{equation} 
The first two terms of \Eq{dpm} recover exactly the Fisher--Jin theory of
short--range wetting \cite{jin93,fisher94}. Since upon performing a lateral
average, the linear term in $\delta\ell$ vanishes, it is clear that the
Fisher--Jin result does not give capillary wave broadening at all, and can
therefore not possibly yield the film--height surface tension arising
from the  external field contribution. That is not a problem in the study of
short--range wetting, where the field is of zero range, but is very relevant in
practical applications discussed here.

A similar expansion performed for \Eq{rhoroughnew} for the case of a liquid--vapor
interface ($h_{\rm wl}(z)=0$) is identical to this equation, except for the $\mp\frac{d
\rho_{\pi}}{d\ell}$ term of \Eq{rhoroughnew}, which here transforms the $\mp$
sign inversion between  branches somewhat more naturally as $(z-\ell_{\pi})
  \frac{d \rho_{\pi}}{d\ell}$. Whence, to the order of approximation that is
implied in \Eq{rhoroughnew}, \Eq{dpm} seems a rather reasonable guess. An advantage
of the heuristic approach is that the crossing criterion is obeyed by
construction for whatever large interfacial displacement. This is not the case
of \Eq{rhoroughnew}, because the boundary condition can only be solved in practice
by performing an expansion in small powers of $\delta\ell$.

The above equation can now be laterally averaged, providing the capillary wave
broadened profile. Since the terms linear in the interface displacement and
Laplacian vanish, we are left with:
\begin{equation}\label{eq:ncwb4}
 \rho(z;\Sigma) = \rho_{\pi}(z) +
 \frac{1}{2} \frac{d^2 \rho_{\pi}}{d\ell^2} \langle \delta\ell^2 \rangle_{\rpar} +
 \frac{1}{2} (z-\ell_{\pi}) \frac{d \rho_{\pi}}{d\ell} \langle (\nabla\ell)^2 \rangle_{\rpar} 
\end{equation} 
a result that is again of very similar form to \Eq{ncwb2}.

The picture that emerges from this discussion is that the addition of
contributions in the gradient and Laplacian
distort the density profile of the planar film
beyond mere interface translations, compressing or relaxing iso--density curves
parallel to the capillary wave perturbation. This is illustrated in
Fig.\ref{fig:sketch_iso},
which shows schematically iso--density curves for the classical
theory, compared with results from $\Eq{dpm}$. In the classical theory,
the  perturbation propagates parallel to the initial wave front
imposed by the crossing criterion. In the modified theory, however, iso--density
lines at one phase (say the vapor phase) detach the original wave front for
perturbations protruding into the liquid,
while they approach the wave front when the  liquid phase protrudes into the
vapor (and likewise for iso-density lines in the liquid phase).
Whence, relative to the classical theory, the perturbation is relaxed faster in
a phase that is receding in favor of the opposite (whence, contributing to a
smaller broadening than implied in the classical theory); but relaxes slower when it is
that phase which protrudes into the opposite (therefore, contributing to a
stronger broadening than predicted by the classical theory).
Subsequent calculations in the framework of the nonlocal theory \cite{fernandez13}
are consistent with our results,  Ref.\cite{macdowell13,macdowell12}.

\begin{figure}
\begin{center}
\includegraphics[clip,scale=0.5,angle=0]{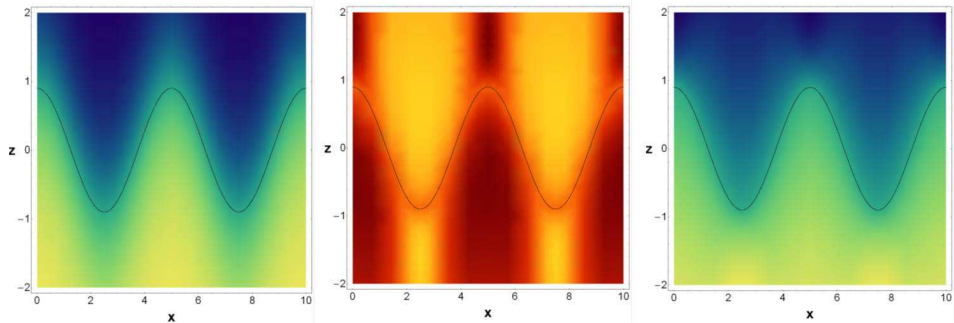} 
\end{center}
\caption{Density plots of a monochromatic capillary perturbation as described by
the classical (left), and the improved capillary wave theories (right). Dark blue
corresponds to the bulk vapor density (top phase), dark yellow to the bulk
liquid density (bottom phase), green values correspond to intermediate densities.
The figure in the center shows differences between the improved and the
classical theories, with yellow indicating positive differences,   red color
negative differences, and orange, no difference. The perturbation is the same as
explained in the caption of Fig.\ref{fig:sketch_iso}.
\label{fig:mapas}
}
\end{figure}

On laterally averaging these two opposing effects, the net contribution is that
of an increased capillary wave broadening, as implied in \Eq{ncwb2} and
\Eq{ncwb4}. Figure \ref{fig:mapas} illustrates this effect for the simple monochromatic
perturbation already considered previously in Fig.\ref{fig:sketch_iso}. The color code which
is darker for densities close to the bulk phase, shows that the relaxation is fast in the
 phase receding from the equimolar surface, but is slow
in the phase protruding into the equimolar surface (this can be best seen by considering the
color code along a horizontal line close to the top or bottom of the figure).
In order to illustrate the difference more clear, the central panel of the
figure  shows the density differences between the improved and the classical theories.
A dominant positive region in the low density phase and an oppositely dominant
negative region in the high density phase indicate enhanced capillary wave
broadening.

\section{Conclusions and outlook}

The most important outcome of this review is an  improved form of the Interfacial
Hamiltonian Model \Eq{ifh} employed in surface thermodynamics for the study of
adsorbed condensates.  According to our recent findings, a better description of
thin films of height $\ell(\rpar)$ subject to the effect of a finite disjoining pressure 
in the small gradient regime is:
\begin{equation}\label{eq:sum1}
H[\ell(\rpar)] = \int  \left \{ g(\ell) + \frac{1}{2}\gamma(\ell)
[ \nabla \ell]^2 \right \} d  \rpar
\end{equation}
where $\gamma(\ell)$ is a film--height dependent surface tension that
asymptotically recovers the result of a free liquid--vapor interface,
$\gamma_{\infty}$. To leading order, it is found that \cite{macdowell13}: 
\begin{equation}\label{eq:sum2}
  \gamma(\ell) = \gamma_{\infty} - \xi_{\infty}^2 \frac{d \Pi(\ell)}{d\ell}
\end{equation} 
where $ \xi_{\infty}$ is the bulk correlation length. In the limit where the
square gradient coefficient is a constant equal to the surface tension, the
extremum of the Interface Hamiltonian provides a generalized or augmented
Young--Laplace equation that is extensively employed to determine the
equilibrium properties of condensates
\cite{philip77,sharma93,degennes85,starov09}. The explicit $\ell$ dependence of
$\gamma(\ell)$, however, shows that the stationary condition of the Interface Hamiltonian
may be more complicated than suggested by the augmented
Young--Laplace equation.

The above result should be accurate at least
in the small curvature regime, which,
for the case of sessile drops may correspond to contact angles as large as 30
degrees \cite{tretyakov13}. The accuracy may be also limited to
a range of film heights where the effect
of an external field is the dominant contribution, as is the case for the
ubiquitous van der Waals forces. It is important to note, however, that this
result does not apply to the immediate vicinity of the substrate, where the role
of short range forces is significant and one can hardly expect  such a
concise $\ell$ dependence for $\gamma(\ell)$ as given by \Eq{sum2}.

The film height dependence of the surface tension close to the substrate has
been recognized for some time \cite{jin93,parry06,bernardino09}, but most
previous studies have been devoted to the analysis of films subject to short--range
forces, which are less relevant in practical applications. The result presented
here is consistent with expectations from a nonlocal theory of interfaces
applied to long range forces \cite{bernardino09}, but is cast here 
in terms of the fluid--substrate disjoining pressure, and is therefore
of more general validity.

The conclusions indicated above result from the study of thermal capillary waves
of adsorbed liquid films \cite{macdowell13}. The Interfacial Hamiltonian  provides a capillary wave
spectrum that depends on the disjoining pressure and surface tension of the
liquid--vapor interface. Studying the fluctuations by means of computer simulations,
we have found that the capillary wave spectrum provides disjoining pressures
that are in full agreement with independent results \cite{gregorio12} obtained from thermodynamic
integration \cite{macdowell11}. However, the study reveals the need to introduce a film height
dependent surface tension closely following expectations from \Eq{sum2}
\cite{macdowell13}. 

The theoretical analysis shows that the height dependence of \Eq{sum2} stems from 
capillary wave perturbations which distort the density profile of the planar film
beyond mere interface translations, compressing or relaxing iso--density curves
parallel to the capillary wave front \cite{macdowell13}. Accordingly, the iso-density curves convey
information on the substrate's external field to terms of order square gradient
and lead to an effective film--height dependent contribution to the surface tension.

Whereas the conclusions embodied in \Eq{sum1} and \Eq{sum2} are possibly the
most relevant, our study has also revealed details of the fluid interface of
adsorbed films previously unnoticed. Particularly, we have shown that the
classical result for capillary wave broadening is modified due to an additional
broadening mechanism of order square gradient in the film fluctuations. An
important test would require to gauge this hypothesis against x-ray
reflectivity studies of adsorbed films that have reported significant discrepancies
with the classical theory \cite{doerr99,wang99,heilmann01,plech02,sferrazza07}.
Of particular interest for experimental verification are systems with large
correlation lengths and small surface tensions \cite{wang99,aarts04}, which, according to \Eq{sum2}
should enhance the film height dependency of $\gamma(\ell)$.

The results of \Eq{sum1} and \Eq{sum2} have many implications in 
the study of adsorption phenomena at distances to the substrate where
$\Pi'(\ell)$ is still relevant.  One obvious application  is the study of
the three phase contact line of droplets \cite{degennes04,starov09}, which
has generated enormous debate in the past. The structure of such condensates
close to a substrate is no longer
dictated by Young's equation. Rather, it is determined from the extremalisation
of an Interfacial Hamiltonian Model \cite{starov09}, which, for the limiting
case of constant surface tension provides the augmented Young--Laplace equation
\cite{philip77,sharma93}.
According to our analysis, a more accurate description of the droplet structure should 
account explicitly for the film--height dependence of the surface tension. The structural changes
which apply in the range  $\ell^{-4}$, should have implications for the study of
the line tension as described by the classical theory \cite{indekeu94}.

Another important application refers to the dewetting dynamics  of adsorbed
liquid films, whether as described by linear theories
\cite{vrij66,ruckenstein74}, or the more involved non--linear treatment
\cite{sharma98,thiele01}. 
In both cases, the interface fluctuations leading to the film rupture are inhibited by the surface
tension. Already at  first sight,  \Eq{sum2} seems to suggest that metastable films, leading to
rupture by nucleation, could be
stabilized relative to the classical expectations, while unstable films
exhibiting spinodal decomposition might be rather, destabilized further.
Another fingerprint of \Eq{sum2} that could be probed in dewetting experiments
is the hole--hole correlations, which, as confirmed by experimental findings
\cite{xie98,seemann01b}, is closely related to the parallel correlation length
$\xi_{\parallel}$. In the modified picture that emerges from this review, the
expected parallel correlation length deviates from the classical value,
$\gamma_{\infty}/g''(\ell)$, by a constant equal to the bulk correlation length
of the adsorbed fluid (c.f. \Eq{newparallel}). Such effect may be again observed in films exhibiting
strong confinement, small surface tensions and large  bulk correlations, as might
be the case of thin  films of polymer mixtures.

Further work is needed to explore these issues in detail.

\paragraph*{Acknowledgements}
We would like to thank E. Sanz for helpful discussions and ongoing
collaborations on this topic.  We acknowledge financial support from grant
FIS2010-22047-C05-05 of the Spanish Ministerio de Economia y Competitividad, and
project PP2009/ESP/1691 (MODELICO) from Comunidad Autonoma de Madrid.




\section*{Bibliography}

\begin{thebibliography}{100}

\bibitem{seemann05}
R. Seemann {\it et~al.}, Proc. Nat. Acad. Sci. {\bf 102},  1848  (2005).

\bibitem{burns96}
M.~A. Burns {\it et~al.}, Proc. Nat. Acad. Sci. {\bf 93},  5556  (1996).

\bibitem{king03}
J.~W. King and L.~L. Williams, Current Op. Solid State Mat. Sci. {\bf 7},  413
  (2003).

\bibitem{cabanas04}
A. Caba{\~n}as, D.~P. Long, and J.~J. Watkins, Chem. Materials {\bf 16},  2028
  (2004).

\bibitem{israelachvili91}
J.~N. Israelachvili, {\em Intermolecular and Surfaces Forces}, 2nd  ed.
  (Academic Press, London, 1991).

\bibitem{milchev01b}
A.~I. Milchev and A.~A. Milchev, Europhys. Lett {\bf 56},  695  (2001).

\bibitem{gretz66}
R.~D. Gretz, J. Chem. Phys. {\bf 45},  3160  (1966).

\bibitem{boruvka77}
L. Boruvka and A.~W. Neumann, J. Chem. Phys. {\bf 66},  5464  (1977).

\bibitem{churaev82}
N. Churaev, V. Starov, and B. Derjaguin, J. Colloid. Interface Sci. {\bf 89},
  16  (1982).

\bibitem{widom95}
B. Widom, J. Chem. Phys. {\bf 99},  2803  (1995).

\bibitem{bresme98}
F. Bresme and N. Quirke, Phys. Rev. Lett. {\bf 80},  3791  (1998).

\bibitem{pompe00}
T. Pompe and S. Herminghaus, Phys. Rev. Lett. {\bf 85},  1930  (2000).

\bibitem{wang01}
J.~Y. Wang, S. Betelu, and B.~M. Law, Phys. Rev. E {\bf 64},  1601  (2001).

\bibitem{pompe02}
T. Pompe, Phys. Rev. Lett. {\bf 89},  076102  (2002).

\bibitem{binder11b}
K. Binder {\it et~al.}, J. Stat. Phys. {\bf 144},  690  (2011).

\bibitem{degennes85}
P.~G. de~Gennes, Rev. Mod. Phys. {\bf 57},  827  (1985).

\bibitem{drelich96}
J. Drelich, Colloids. Surf. A {\bf 116},  43  (1996).

\bibitem{amirfazli04}
A. Amirfazli and A. Neumann, Adv. Colloid Interface Sci. {\bf 110},  121
  (2004).

\bibitem{velarde11}
{\em European Physical Journal--Special Topics}, edited by M.~G. Velarde
  (Springer, Berlin, 2011), Vol.~197.

\bibitem{philip77}
J.~R. Philip, J. Chem. Phys. {\bf 66},  5069  (1977).

\bibitem{robbins91}
M.~O. Robbins, D. Andelman, and J.-F. Joanny, Phys. Rev. A {\bf 43},  4344
  (1991).

\bibitem{sharma93}
A. Sharma, Langmuir {\bf 9},  3580  (1993).

\bibitem{starov09}
V.~M. Starov and M.~G. Velarde, J. Phys.: Condens. Matter {\bf {21}},  464121
  ({2009}).

\bibitem{derjaguin92}
B. Derjaguin and N. Churaev, Progress in Surface Science {\bf 40},  272
  (1992).

\bibitem{derjaguin92b}
B. Derjaguin, Progress in Surface Science {\bf 40},  254  (1992).

\bibitem{dietrich88}
S. Dietrich,  in {\em Phase Transitions and Critical Phenomena}, edited by C.
  Domb and J.~L. Lebowitz (Academic, New York, 1988), Vol.~12, pp.\ 1--89.

\bibitem{schick90}
M. Schick, {\em Liquids at Interfaces}, {\em Les Houches Lecture Notes}
  (Elsevier Science Publishers, Amsterdam, 1990), pp.\ 1--89.

\bibitem{buff65}
F.~P. Buff, R.~A. Lovett, and F.~H. Stillinger, Phys. Rev. Lett. {\bf 15},  621
   (1965).

\bibitem{fisher85}
D.~S. Fisher and D.~A. Huse, Phys. Rev. B {\bf 32},  247  (1985).

\bibitem{degennes04}
P.~G. de~Gennes, F. Brochard-Wyart, and D. Qu{\'e}r{\'e}, {\em Capillarity and
  Wetting Phenomena} (Springer, New York, 2004).

\bibitem{dobbs93}
H.~T. Dobbs and J.~O. Indekeu, Physica. A {\bf 201},  457  (1993).

\bibitem{bauer99b}
C. Bauer, S. Dietrich, and A.~O. Parry, Europhys. Lett {\bf 47},  474  (1999).

\bibitem{vrij66}
A. Vrij, Discuss. Faraday Soc. {\bf 42},  23  (1966).

\bibitem{safran94}
S.~A. Safran, {\em Statistical Thermodynamics of Surfaces, Interfaces and
  Membranes}, 1st  ed. (Addison-Wesley, Reading, 1994).

\bibitem{schick92}
M. Schick and P. Taborek, Phys. Rev. B {\bf 46},  7312  (1992).

\bibitem{binder88}
K. Binder and D.~P. Landau, Phys. Rev. B {\bf 37},  1745  (1988).

\bibitem{jin93}
A.~J. Jin and M.~E. Fisher, Phys. Rev. B {\bf 47},  7365  (1993).

\bibitem{fisher94}
M.~E. Fisher, A.~J. Jin, and A.~O. Parry, Ber. Bunsenges. Phys. Chem. {\bf 98},
   357  (1994).

\bibitem{parry06}
A.~O. Parry, C. Rasc{\'o}n, N.~R. Bernardino, and J.~M. Romero-Enrique, J.
  Phys.: Condens. Matter {\bf 18},  6433  (2006).

\bibitem{parry07}
A.~O. Parry, C. Rasc{\'o}n, N.~R. Bernardino, and J.~M. Romero-Enrique, J.
  Phys.: Condens. Matter {\bf 19},  416105  (2007).

\bibitem{ragil96}
K. Ragil {\it et~al.}, Phys. Rev. Lett. {\bf 77},  1532  (1996).

\bibitem{shahidzadeh98}
N. Shahidzadeh {\it et~al.}, Phys. Rev. Lett. {\bf 80},  3992  (1998).

\bibitem{bernardino08}
N.~R. Bernardino, Ph.D. thesis, Imperial College, 2008.

\bibitem{bernardino09}
N.~R. Bernardino, A.~O. Parry, C. Rasc{\'o}n, and J.~M. Romero-Enrique, J.
  Phys.: Condens. Matter {\bf 21},  465105  (2009).

\bibitem{macdowell13}
L.~G. MacDowell, J. Benet, and N.~A. Katcho, Phys. Rev. Lett. {\bf 111},
  047802  (2013).

\bibitem{rowlinson82b}
J. Rowlinson and B. Widom, {\em Molecular Theory of Capillarity} (Clarendon,
  Oxford, 1982).

\bibitem{tidswell91}
I.~M. Tidswell, T.~A. Rabedeau, P.~S. Pershan, and S.~D. Kosowsky, Phys. Rev.
  Lett. {\bf 66},  2108  (1991).

\bibitem{doerr99}
A.~K. Doerr {\it et~al.}, Phys. Rev. Lett. {\bf 83},  3470  (1999).

\bibitem{mora03}
S. Mora {\it et~al.}, Phys. Rev. Lett. {\bf 90},  216101  (2003).

\bibitem{pang11}
L. Pang, D.~P. Landau, and K. Binder, Phys. Rev. Lett. {\bf 106},  236102
  (2011).

\bibitem{fernandez12}
E.~M. Fern{\'a}ndez, E. Chac{\'o}n, and P. Tarazona, Phys. Rev. B {\bf 86},
  085401  (2012).

\bibitem{macdowell11}
L.~G. MacDowell, Euro. Phys. J. ST {\bf 197},  131  (2011).

\bibitem{gregorio12}
R. de~Gregorio {\it et~al.}, J. Chem. Phys. {\bf 136},  104703  (2012).

\bibitem{evans92}
R. Evans,  in {\em Fundamentals of Inhomogenous Fluids}, edited by D. Henderson
  (Marcel Dekker, New York, 1992), Chap.~3, pp.\ 85--175.

\bibitem{henderson92}
D. Henderson, {\em Fundamentals of Inhomogenous Fluids} (Marcel Dekker, New
  York, 1992).

\bibitem{hansen86}
J.-P. Hansen and I.~R. McDonald, {\em Theory of Simple Liquids} (Academic
  Press, London, 1986).

\bibitem{mcquarrie76}
D.~A. McQuarrie, {\em Statistical Mechanics} (Harper \& Row, New York, 1976).

\bibitem{tang04}
Y. Tang and J. Wu, Phys. Rev. E {\bf 70},  011201  (2004).

\bibitem{tang05}
Y. Tang, J. Chem. Phys. {\bf 123},  204704  (2005).

\bibitem{tang07}
Y. Tang, J. Chem. Phys. {\bf 126},  249901  (2007).

\bibitem{barker82}
J.~A. Barker and J.~R. Henderson, J. Chem. Phys. {\bf 76},  6303  (1982).

\bibitem{cahn58}
J.~W. Cahn and J.~E. Hilliard, J. Chem. Phys. {\bf 28},  258  (1958).

\bibitem{parry04}
A.~O. Parry, J.~M. Romero-Enrique, and A. Lazarides, Phys. Rev. Lett. {\bf 93},
   086104  (2004).

\bibitem{palanco13}
J.~M.~G. Palanco, Ph.D. thesis, Universidad Complutense de Madrid, 2013.

\bibitem{iwamatsu93}
M. Iwamatsu, J. Phys.: Condens. Matter {\bf 5},  7537  (1993).

\bibitem{iwamatsu94}
M. Iwamatsu, J. Phys.: Condens. Matter {\bf 6},  L173  (1994).

\bibitem{bykov99b}
T.~V. Bykov and X.~C. Zeng, J. Chem. Phys. {\bf 111},  10602  (1999).

\bibitem{hemingway81}
S.~J. Hemingway, J.~R. Henderson, and J.~S. Rowlinson, Faraday Symp. Chem. Soc.
  {\bf 16},  33  (1981).

\bibitem{ebner77}
C. Ebner and W.~F. Saam, Phys. Rev. Lett. {\bf 38},  1486  (1977).

\bibitem{horn81}
R.~G. Horn and J.~N. Israelachvili, J. Chem. Phys. {\bf 75},  1400  (1981).

\bibitem{chernov88}
A.~A. Chernov and L.~V. Mikheev, Phys. Rev. Lett. {\bf 60},  2488  (1988).

\bibitem{tarazona83}
P. Tarazona and R. Evans, Mol. Phys. {\bf 48},  799  (1983).

\bibitem{tarazona85}
P. Tarazona and .~L. Vicente, Mol. Phys. {\bf 56},  557  (1985).

\bibitem{henderson05}
J.~R. Henderson, Phys. Rev. E {\bf 72},  051602  (2005).

\bibitem{klapp08}
S.~H.~L. Klapp, S. Grandner, Y. Zeng, and R. von Klitzing, J. Phys.: Condens.
  Matter {\bf 20},  494232  (2008).

\bibitem{tarazona84}
P. Tarazona, Mol. Phys. {\bf 52},  81  (1984).

\bibitem{evans94}
R. Evans, R.~J. F.~L. de~Carvalho, J.~R. Henderson, and D.~C. Hoyle, J. Chem.
  Phys. {\bf 100},  591  (1994).

\bibitem{henderson94}
J.~R. Henderson, Phys. Rev. E {\bf 50},  4836  (1994).

\bibitem{evans09}
R. Evans and J.~R. Henderson, J. Phys.: Condens. Matter {\bf 21},  474220
  (2009).

\bibitem{evans93b}
R. Evans {\it et~al.}, Mol. Phys. {\bf 80},  755  (1993).

\bibitem{klapp08b}
S.~H.~L. Klapp, Y. Zeng, D. Qu, and R. von Klitzing, Phys. Rev. Lett. {\bf
  100},  118303  (2008).

\bibitem{evans86b}
R. Evans and U.~M.~B. Marconi, Phys. Rev. A {\bf 34},  3504  (1986).

\bibitem{benet11}
J. Benet, Master's thesis, Universidad Complutense de Madrid, 2011.

\bibitem{dietrich91}
S. Dietrich and M. Napi\'orkowski, Phys. Rev. A {\bf 43},  1861  (1991).

\bibitem{mecke99b}
K.~R. Mecke and S. Dietrich, Phys. Rev. E {\bf 59},  6766  (1999).

\bibitem{werner99}
A. Werner, M. Muller, F. Schmid, and K. Binder, J. Chem. Phys. {\bf 110},  1221
   (1999).

\bibitem{sferrazza07}
M. Sferrazza and C. Carelli, J. Phys.: Condens. Matter {\bf 19},  073102
  (2007).

\bibitem{goldenfeld92}
N. Goldenfeld, {\em Lectures on Phase Transitions and the Renormalization
  Group} (Perseus Books, Reading, Massachusetts, 1992).

\bibitem{kayser86}
R.~F. Kayser, Phys. Rev. A {\bf 33},  1948  (1986).

\bibitem{mueller96}
M. M{\"u}ller and M. Schick, J. Chem. Phys. {\bf 105},  8282  (1996).

\bibitem{mueller96b}
M. M{\"u}ller and M. Schick, J. Chem. Phys. {\bf 105},  8885  (1996).

\bibitem{mueller00}
M. M{\"u}ller and L.~G. MacDowell, Macromolecules {\bf 33},  3902  (2000).

\bibitem{milchev02}
A. Milchev and K. Binder, Europhys. Lett {\bf 59},  81  (2002).

\bibitem{davidchack06}
R.~L. Davidchack, J.~R. Morris, and B.~B. Laird, J. Chem. Phys. {\bf 125},
  094710  (2006).

\bibitem{zykova-tilman10}
T. Zykova-Tilman, J. Horbach, and K. Binder, J. Chem. Phys. {\bf 133},  014705
  (2010).

\bibitem{rozas11}
R.~E. Rozas and J. Horbach, Europhys. Lett {\bf 93},  26006  (2011).

\bibitem{mecke01}
K.~R. Mecke, J. Phys.: Condens. Matter {\bf 13},  4615  (2001).

\bibitem{gelfand90}
M.~P. Gelfand and M.~E. Fisher, Physica. A {\bf 166},  1  (1990).

\bibitem{weeks77}
J.~D. Weeks, J. Chem. Phys. {\bf 67},  3106  (1977).

\bibitem{abraham81}
D.~B. Abraham, Phys. Rev. Lett. {\bf 47},  545  (1981).

\bibitem{vink05}
R.~L.~C. Vink, J. Horbach, and K. Binder, J. Chem. Phys. {\bf 122},  134905
  (2005).

\bibitem{jasnow84}
D. Jasnow, Rep. Prog. Phys. {\bf 47},  1059  (1984).

\bibitem{binder82}
K. Binder, Phys. Rev. A {\bf 25},  1699  (1982).

\bibitem{chen95}
L.-J. Chen, J. Chem. Phys. {\bf 103},  10214  (1995).

\bibitem{aguado01b}
A. Aguado, W. Scott, and P.~A. Madden, J. Chem. Phys. {\bf 115},  8612  (2001).

\bibitem{tarazona07}
P. Tarazona, R. Checa, and E. Chacon, Phys. Rev. Lett. {\bf 99},  196101
  (2007).

\bibitem{paulus08}
M. Paulus, C. Gutt, and M. Tolan, Phys. Rev. B {\bf 78},  235419  (2008).

\bibitem{pershan12}
P.~S. Pershan and M. Schlossman, {\em Liquid Surfaces and Interfaces:
  Synchrotron {X-ray} Methods} (Cambridge University Press, Cambridge, 2012).

\bibitem{blokhuis09}
E.~M. Blokhuis, J. Chem. Phys. {\bf 130},  074701  (2009).

\bibitem{daillant00}
J. Daillant and M. Alba, Rep. Prog. Phys. {\bf 63},  1725  (2000).

\bibitem{schwartz90}
D.~K. Schwartz {\it et~al.}, Phys. Rev. A {\bf 41},  5687  (1990).

\bibitem{sanyal91}
M.~K. Sanyal, S.~K. Sinha, K.~G. Huang, and B.~M. Ocko, Phys. Rev. Lett. {\bf
  66},  628  (1991).

\bibitem{ocko94}
B.~M. Ocko {\it et~al.}, Phys. Rev. Lett. {\bf 72},  242  (1994).

\bibitem{tidswell91b}
I.~M. Tidswell {\it et~al.}, Phys. Rev. B {\bf 44},  10869  (1991).

\bibitem{heilmann01}
R.~K. Heilmann, M. Fukuto, and P.~S. Pershan, Phys. Rev. B {\bf 63},  205405
  (2001).

\bibitem{plech02}
A. Plech {\it et~al.}, Phys. Rev. E {\bf 65},  061604  (2002).

\bibitem{sengers89}
J.~V. Sengers and J.~M.~J. van Leeuwen, Phys. Rev. A {\bf 39},  6346  (1989).

\bibitem{lacasse98}
M.-D. Lacasse, G.~S. Grest, and A.~J. Levine, Phys. Rev. Lett. {\bf 80},  309
  (1998).

\bibitem{sides99}
S.~W. Sides, G.~S. Grest, and M.-D. Lacasse, Phys. Rev. E {\bf 60},  6708
  (1999).

\bibitem{geysermans10}
P. Geysermans and V. Pontikis, J. Chem. Phys. {\bf 133},  074706  (2010).

\bibitem{carelli05}
C. Carelli {\it et~al.}, Phys. Rev. E {\bf 72},  031807  (2005).

\bibitem{luo06}
G. Luo {\it et~al.}, J. Electroanalytical Chem. {\bf 593},  142  (2006).

\bibitem{werner97}
A. Werner, F. Schmid, M. Muller, and K. Binder, J. Chem. Phys. {\bf 107},  8175
   (1997).

\bibitem{sferrazza97}
M. Sferrazza {\it et~al.}, Phys. Rev. Lett. {\bf 78},  3693  (1997).

\bibitem{parry92}
A.~O. Parry and R. Evans, Physica. A {\bf 181},  250  (1992).

\bibitem{kerle96}
T. Kerle, J. Klein, and K. Binder, Phys. Rev. Lett. {\bf 77},  1318  (1996).

\bibitem{luo06b}
G. Luo {\it et~al.}, J. Phys. Chem. B {\bf 110},  4527  (2006).

\bibitem{hou13}
B. Hou {\it et~al.}, J. Phys. Chem. B {\bf 117},  5365  (2013).

\bibitem{jeng98}
U.-S. Jeng, L. Esibov, L. Crow, and A. Steyerl, J. Phys.: Condens. Matter {\bf
  10},  4955  (1998).

\bibitem{bedeaux85b}
D. Bedeaux and J.~D. Weeks, J. Chem. Phys. {\bf 82},  972  (1985).

\bibitem{benjamin92}
I. Benjamin, J. Chem. Phys. {\bf 97},  1432  (1992).

\bibitem{benet13}
J. Benet, J.~M.~G. Palanco, E. Sanz, and L.~G. MacDowell, in preparation.
  (unpublished).

\bibitem{fisher82}
M.~P.~A. Fisher, D.~S. Fisher, and J.~D. Weeks, Phys. Rev. Lett. {\bf 48},  368
   (1982).

\bibitem{mitrinovic00}
D.~M. Mitrinovi\ifmmode~\acute{c}\else \'{c}\fi{} {\it et~al.}, Phys. Rev.
  Lett. {\bf 85},  582  (2000).

\bibitem{sikkenk87}
J.~H. Sikkenk, J.~O. Indekeu, J.~M.~J. van Leeuwen, and E.~O. Vossnack, Phys.
  Rev. Lett. {\bf 59},  98  (1987).

\bibitem{finn88}
J.~E. Finn and P.~A. Monson, Mol. Phys. {\bf 65},  1345  (1988).

\bibitem{finn89}
J.~E. Finn and P.~A. Monson, Phys. Rev. A {\bf 39},  6402  (1989).

\bibitem{sokolowski90}
S. Sokolowski and J. Fischer, Phys. Rev. A {\bf 41},  6866  (1990).

\bibitem{rutledge92}
J.~E. Rutledge and P. Taborek, Phys. Rev. Lett. {\bf 69},  937  (1992).

\bibitem{cheng93}
E. Cheng {\it et~al.}, Phys. Rev. Lett. {\bf 70},  1854  (1993).

\bibitem{chacon05}
E. Chacon and P. Tarazona, J. Phys.: Condens. Matter {\bf 17},  S3493  (2005).

\bibitem{usabiaga09}
F.~B. Usabiaga and D. Duque, Phys. Rev. E {\bf 79},  046709  (2009).

\bibitem{chacon03}
E. Chacon and P. Tarazona, Phys. Rev. Lett. {\bf 91},  166103  (2003).

\bibitem{chacon09}
E. Chac\'on {\it et~al.}, Phys. Rev. B {\bf 80},  195403  (2009).

\bibitem{bresme08}
F. Bresme, E. Chacon, P. Tarazona, and K. Tay, Phys. Rev. Lett. {\bf 101},
  056102  (2008).

\bibitem{fernandez11b}
E.~M. Fern{\'a}ndez, E. Chac{\'o}n, and P. Tarazona, Phys. Rev. E {\bf 84},
  205435  (2011).

\bibitem{blake72}
T.~D. Blake and J.~A. Kitchener, J. Chem. Soc., Faraday Trans. {\bf 68},  1435
  (1972).

\bibitem{blake75}
T.~D. Blake, J. Chem. Soc., Faraday Trans. 1 {\bf 71},  192  (1975).

\bibitem{vazquez05}
R. Vazquez {\it et~al.}, J. Colloid. Interface Sci. {\bf 284},  652  (2005).

\bibitem{scheludko67}
A. Scheludko, Adv. Colloid Interface Sci. {\bf 1},  391  (1967).

\bibitem{langevin11}
D. Langevin, C. Marquez-Beltran, and J. Delacotte, Adv. Colloid Interface Sci.
  {\bf 168},  124  (2011).

\bibitem{seemann01b}
R. Seemann, S. Herminghaus, and K. Jacobs, Phys. Rev. Lett. {\bf 86},  5534
  (2001).

\bibitem{kim99}
H.~I. Kim, C.~M. Mate, K.~A. Hannibal, and S.~S. Perry, Phys. Rev. Lett. {\bf
  82},  3496  (1999).

\bibitem{macdowell05}
L.~G. MacDowell and M. M{\"u}ller, J. Phys.: Condens. Matter {\bf 17},  S3523
  (2005).

\bibitem{macdowell06}
L.~G. MacDowell and M. M{\"u}ller, J. Chem. Phys. {\bf 124},  084907  (2006).

\bibitem{berg92}
B.~A. Berg and T. Neuhaus, Phys. Rev. Lett. {\bf 68},  9  (1992).

\bibitem{fitzgerald99}
M. Fitzgerald, R.~R. Picard, and R.~N. Silver, Europhysics Lett. {\bf 46},  282
   (1999).

\bibitem{errington04}
J.~R. Errington, Langmuir {\bf 20},  3798  (2004).

\bibitem{shen05}
V.~K. Shen and J.~R. Errington, J. Chem. Phys. {\bf 122},  064508  (2005).

\bibitem{grzelak08}
E.~M. Grzelak and J.~R. Errington, J. Chem. Phys. {\bf 128},  014710  (2008).

\bibitem{kumar11}
V. Kumar, S. Sridhar, and J.~R. Errington, J. Chem. Phys. {\bf 135},  184702
  (2011).

\bibitem{rane11}
K.~S. Rane, V. Kumar, and J.~R. Errington, J. Chem. Phys. {\bf 135},  234102
  (2011).

\bibitem{shi02}
W. Shi, X. Zhao, and J.~K. Johnson, Mol. Phys. {\bf 100},  2139  (2002).

\bibitem{tolstov62}
G.~P. Tolstov, {\em Fourier Series} (Dover, New York, 1962).

\bibitem{vanleeuwen89}
J.~V. Leeuwen and J. Sengers, Physica. A {\bf 157},  839  (1989).

\bibitem{wang99}
J. Wang {\it et~al.}, Phys. Rev. Lett. {\bf 83},  564  (1999).

\bibitem{macdowell12}
L.~G. MacDowell, J. Benet, and N.~A. Katcho, Capillary fluctuations and
  interface potential of an adsorbed liquid film, unpublished comunication sent
  to P. Tarazona, E. Chacon and E. M. Fernandez, 2012.

\bibitem{fernandez13}
E.~M. Fern\'andez {\it et~al.}, Phys. Rev. Lett. {\bf 111},  096104  (2013),
  c.f. Ref.\cite{macdowell13,macdowell12}.

\bibitem{tretyakov13}
N. Tretyakov, M. Muller, D. Todorova, and U. Thiele, J. Chem. Phys. {\bf 138},
  064905  (2013).

\bibitem{aarts04}
D.~G. Aarts, M. Schmidtt, and H.~N.~K. Lekkerkerker, Science {\bf 304},  847
  (2004).

\bibitem{indekeu94}
J.~O. Indekeu, Int. J. Mod.Phys.B {\bf 8},  309  (1994).

\bibitem{ruckenstein74}
E. Ruckenstein and R.~K. Jain, J. Chem. Soc.{,} Faraday Trans. 2 {\bf 70},  132
   (1974).

\bibitem{sharma98}
A. Sharma and R. Khanna, Phys. Rev. Lett. {\bf 81},  3463  (1998).

\bibitem{thiele01}
U. Thiele, M.~G. Velarde, and K. Neuffer, Phys. Rev. Lett. {\bf 87},  016104
  (2001).

\bibitem{xie98}
R. Xie {\it et~al.}, Phys. Rev. Lett. {\bf 81},  1251  (1998).

\end{thebibliography}






\end{document}